\documentclass{aa}  

\usepackage{comment}
\usepackage{graphicx}
\usepackage[dvipsnames]{xcolor}
\usepackage{subcaption}
\graphicspath{ {./figures/} }

\usepackage{txfonts}
\usepackage[%
  breaklinks = true,
  colorlinks = true,
  urlcolor   = blue,
  citecolor  = blue,
  linkcolor  = blue,
]{hyperref}
\usepackage{soul}
\usepackage{nameref}
\usepackage{makecell}
\usepackage{booktabs}
\usepackage[flushleft]{threeparttable}
\usepackage[labelfont=bf]{caption}
\usepackage{hhline}
\usepackage[utf8]{inputenc}
\newcommand{\Ionline}[3]{\ion{#1}{#2} #3~{\AA}}
\newcommand{\ionline}[3]{\ion{#1}{#2} #3~{\AA}}

\newcommand{\secondpaperref}{2024A&A...683A..95F}

\defcitealias{2023A&A...675A..97F}{F2023}
\defcitealias{\secondpaperref}{F2024}

\begin{document} 

   \title{Extreme-ultraviolet (EUV) observables of simulated plasmoid-mediated reconnection in the solar corona}
   \titlerunning{EUV observables of simulated plasmoid-mediated reconnection in the solar corona}
   \authorrunning{Ø.H.Færder et al.}

   \author{Ø. H. Færder\inst{1,2},
          D. Nóbrega-Siverio\inst{3,4,1,2},
          M. Carlsson\inst{1,2}, and
          J. Mart\'inez-Sykora\inst{5,6,1,2}
          }

   \institute{Rosseland Centre for Solar Physics, University of Oslo,
              PO Box 1029, Blindern, NO-0315 Oslo, Norway\\
              \email{o.h.farder@astro.uio.no}
         \and
              Institute of Theoretical Astrophysics, University of Oslo,
              PO Box 1029, Blindern, NO-0315 Oslo, Norway
         \and
             Instituto de Astrof\'isica de Canarias,    E-38205 La Laguna,  Tenerife, Spain
        \and
            Universidad de La Laguna, Dept. Astrof\'isica, E-38206 La  Laguna, Tenerife, Spain
        \and
            Lockheed Martin Solar \& Astrophysics Laboratory, 3251 Hanover Street, Palo Alto, CA 94304, USA
        \and
            Bay Area Environmental Research Institute, NASA Research Park, Moffett Field, CA 94035, USA
            }

% \abstract{}{}{}{}{} 
% 5 {} token are mandatory
 
  \abstract
   {Understanding the role of magnetic reconnection in the heating and dynamics of the solar atmosphere requires detailed observational data of any observable aspect of the reconnection process, including small-scale features such as plasmoids.}
   {Here, we examine the capability of active and upcoming instruments to detect plasmoids generated by reconnection in the corona including low-density regimes.}
  {We used the Bifrost code to perform simulations of plasmoid-mediated reconnection in the corona with a 2D idealized setup: a fan-spine topology with uniform density including thermal conduction. Through a forward-modeling of extreme-ultraviolet (EUV) observables, we checked whether our simulated plasmoids could be detected with the instruments of Solar Dynamics Observatory (SDO) and Solar Orbiter (SO), as well as the upcoming Multi-Slit Solar Explorer (MUSE) and Solar-C missions.}
   {Short-lived ($\sim 10-20$ s) small-scale ($\sim 0.2-0.5$ Mm) coronal plasmoids are not resolvable with the Atmospheric Imaging Assembly (AIA) on board SDO. In contrast, they could be captured with the EUV High-Resolution Imager at the Extreme Ultraviolet Imager (EUI-HRI$_\mathrm{EUV}$) of SO. The spatial and temporal high-resolution planned for the MUSE spectrograph (SG) is adequate to obtain full spectral information of these  plasmoids. To achieve a sufficient signal-to-noise ratio (S/N) for $\sim$0.8 MK plasmoids in the MUSE/SG 171 \AA\ channel, full-raster images are attainable for regions with electron densities above $10^9\ \mathrm{cm^{-3}}$, while sit-and-stare observations are recommended for lower-density regions. The future Solar-C mission could also capture these coronal plasmoids using the EUV High-Throughput Spectroscopic Telescope (EUVST), considering rapid changes in Doppler shift and line widths in different EUV lines caused by plasmoid motions along the current sheet.}
  {With the combined spectra of MUSE/SG and Solar-C/EUVST in multiple emission lines, along with high-resolution images from SO/EUI-HRI$_\mathrm{EUV}$ and MUSE/CI, it should be possible to gain new insights about plasmoid formation in the corona.}

   \keywords{
                magnetohydrodynamics (MHD) --
                magnetic reconnection --
                methods: numerical -- Sun: atmosphere -- Sun: corona -- Sun: magnetic fields
               }

   \maketitle
   
%
%-----------------------------------------------------------------------

\section{Introduction}
\label{sec:intro}

Magnetic reconnection stands out as a candidate for solving the coronal heating problem \citep{1973SoPh...32...81V, 1984A&A...137...63H, 1988ApJ...330..474P}. Numerical simulations have shown the ability of this process to generate a wide range of observable solar phenomena such as Ellerman bombs (EBs), ultraviolet (UV) bursts, coronal bright points (CBPs), surges, coronal jets, and flares  \citep{1995Natur.375...42Y, 1996PASJ...48..353Y, 2001ApJ...549.1160Y, 2016ApJ...822...18N, 2017ApJ...850..153N, 2023ApJ...958L..38N, 2016ApJ...827....4W, 2017Natur.544..452W, 2017ApJ...839...22H, 2019A&A...626A..33H, 2017A&A...601A.122D, 2019A&A...628A...8P, 2021A&A...646A..88N, 2022ApJ...935L..21N, 2023ApJ...955..105R}. Furthermore, numerical simulations demonstrate the connection between magnetic reconnection and the formation of small features within current sheets, such as plasmoids \citep{2017ApJ...841...27N, 2017ApJ...850..153N, 2019A&A...626A..33H, 2019A&A...628A...8P, 2020ApJ...901..148G, 2022A&A...665A.116N, 2023RAA....23c5006L}.

Plasmoids appear in any current sheet due to resistive tearing instability \citep{1963PhFl....6..459F} as long as the Lundquist number is sufficiently high \citep[$S_{\rm L}>10^4$,][]{2007PhPl...14j0703L} and the current sheet width is below a critical value \citep[$a/L<S_{\rm L}^{-1/3}$,][]{2014ApJ...780L..19P}. Plasmoid instability causes the characteristics of the reconnection process, including the reconnection rate, to be significantly different from what predicted with the Sweet-Parker \citep{1958IAUS....6..123S, 1958NCim....8S.188S, 1963ApJS....8..177P} or the Petschek \citep{1964NASSP..50..425P} model, which both assume steady reconnection. In particular, on the Sun, they have been observationally reported related to EBs, UV bursts, surges, coronal jets, and flares \citep[see, e.g.,][]{2017ApJ...851L...6R, 2023A&A...673A..11R, 2019ApJ...885L..15K, 2020ApJ...901..148G, 2022NatCo..13..640Y}. 

In order to broaden our understanding about magnetic reconnection, further explorations of plasmoids are necessary. This could allow us not only to improve our theoretical models about plasmoid-mediated reconnection in the solar atmosphere, but also to establish constraints on the observational detection capabilities of such features using current and future missions. For instance, the Solar Dynamics Observatory \citep[SDO,][]{2012SoPh..275....3P}, with its Atmospheric Imaging Assembly \citep[AIA,][]{ 2012SoPh..275...17L, 2012SoPh..275...41B}, observes the corona through several EUV narrow-band filters, including one focused on the \ion{Fe}{IX} 171 \AA\ line. 
Though AIA has been capable of detecting plasmoids of sizes $\gtrsim 1$ Mm in coronal mass ejections \citep{2019SciA....5.7004G}, flares \citep{2012ApJ...745L...6T, 2022NatCo..13..640Y}, and nanojets \citep{2021NatAs...5...54A}, its moderate spatial resolution of $1\farcs5$ could make it challenging for studying small-scale ($<1$ Mm) plasmoids. 
%plasmoids and other small-scale ($<1$ Mm) features.
In contrast, the recently launched Solar Orbiter (SO) mission, through the Extreme Ultraviolet Imager instrument and its EUV High-Resolution Imager telescope \citep[EUI-HRI$_\mathrm{EUV}$,][]{2020A&A...642A...8R}, can offer coronal images of appreciably improved spatial and temporal resolution focused on the \ion{Fe}{X} 174 \AA\ line, especially at its perihelion distance of 0.3 AU \citep{2023A&A...675A.110B}. 
The near-future launches of the Multi-slit Solar Explorer \citep[MUSE,][]{2019ApJ...882...13C, 2022ApJ...926...53C, 2020ApJ...888....3D, 2022ApJ...926...52D} and the Solar-C \citep{2021SPIE11852E..3KS} missions, are expected to open up more possibilities to delve further into the dynamics of the coronal plasma. In particular, MUSE, with its 35-slit spectrograph (SG), is designed to obtain solar spectra in four bright EUV lines, including \ion{Fe}{IX} 171 \AA, allowing us to study coronal plasma in all temperature ranges  from 0.8 MK to 12 MK on unprecedented timescales. With a spatial pixel size of $0\farcs167\times 0\farcs4$, it will provide an excellent opportunity for observing small-scale plasmoids in the corona. In addition, MUSE will have two context imagers (CI) set to observe the Sun with filters centered in the \ion{He}{II} 304 \AA\ and \ion{Fe}{XII} 195 \AA\ lines. Solar-C is planned to have a better temperature coverage thanks to the single-slit EUV High-Throughput Spectroscopic Telescope \citep[EUVST,][]{2021SPIE11852E..3KS}, which could scan a wide range of EUV wavelengths, including well-known coronal lines such as \ion{Fe}{IX} 171 \AA, \ion{Fe}{X} 174 \AA, and \ion{Fe}{XII} 195 \AA, among many others. Despite having only one slit, in contrast to MUSE, this spectrograph may provide detailed temperature and density diagnostics in the chosen location where the slit is aimed, with a slightly higher spatial ($0\farcs 16$) and spectral resolution than MUSE. Hence, these two upcoming spectrographs complement each other and can together be used to retrieve detailed information on relatively small structures in the corona.

Preparing for observational studies of plasmoids in the solar corona involves understanding the detectable signatures they may leave. In our previous paper \citep[hereafter \citetalias{\secondpaperref}]{\secondpaperref}, we carried out numerical simulations of reconnection in a fan-spine magnetic topology within a 2D coronal domain, showing how different resistivity models produced varying plasmoid characteristics. Therefore, characterising observed plasmoid properties could serve as an indicator of the most suitable resistivity model for reproducing coronal-like reconnection. In this paper, we use simulations akin to those used in the paper by \citetalias{\secondpaperref}  as a foundation for spectral synthesis, aiming at determining the capability of each of the aforementioned space-borne instruments in detecting small ($\sim 0.2-0.5\ \mathrm{Mm}$), short-lived ($\sim 10-20\ \mathrm{s}$) plasmoids in coronal regions with $\sim 0.8\ \mathrm{MK}$.
The structure of the paper is as follows. Section \ref{sec:methods} describes the numerical model along with the methods applied for spectral synthesis. Section \ref{sec:results} gives a brief overview of the simulations and presents the results from the spectral synthesis. Finally, Sect.~\ref{sec:discussion} gives a brief discussion of these results and summarises the conclusions.
%-----------------------------------------------------------------------
\section{Methods}
\label{sec:methods}

In the following sections, we summarize briefly how we set up our numerical simulations and give some details on how we performed our forward-modeling of each synthetic observable studied in this paper.

\subsection{Numerical simulations}
\label{sec:simulations}

The numerical simulations were performed with the Bifrost code \citep{2011A&A...531A.154G}. This code solves the equations of magnetohydrodynamics (MHD) on a 3D Cartesian grid, applying a sixth-order operator for spatial derivatives and a third-order scheme for time derivatives, thereby minimising the numerical diffusion due to the discretization. 
In our special case, we focused on a purely in-plane 2D coronal domain, using an idealized setup where we included Joule heating, viscous heating, and Spitzer conductivity, but gravity is neglected and the equation-of-state is approximated by an ideal gas law with a mean molecular weight of 0.616. We also neglected optically thin losses because including them in an idealized 2D setup like this one requires incorporating an ad hoc heating term to prevent the draining of internal energy, which is difficult to do in a realistic way. This assumption is discussed further in Sect.~\ref{sec:discussion} based on our results.

We performed two simulations using the same initial condition as in the \citetalias{\secondpaperref} paper, namely, a fan-spine magnetic topology given by 
\begin{align}
    B_x &= B_1 e^{-kz} \sin (kx) \label{eq:init-bx}, \\
    B_z &= B_0 + B_1 e^{-kz}  \cos (kx) \label{eq:init-bz}, 
\end{align}
with $k = \pi/16\ \mathrm{Mm^{-1}}$, imposed on a 2D coronal domain with an initial uniform temperature of 0.61 MK and different input values for $B_0$, $B_1$, and the initial mass density $\rho_0$. The first simulation case, which is case G6 from the \citetalias{\secondpaperref} paper, 
has $\rho_0 = 3.0 \times 10^{-16}\ \mathrm{g\ cm^{-3}}$, $B_0 = 3$ G, and $B_1 = 10$ G, 
while the second one, referred to as case G6b, has $\rho_0 = 2.7 \times 10^{-15}\ \mathrm{g\ cm^{-3}}$, $B_0 = 9$ G, and $B_1 = 30$ G. With this, cases G6 and G6b have the same initial plasma-$\beta$ (inflow value $\sim 2$), Alfvén speed (inflow value $\sim 125\ \mathrm{km\ s^{-1}}$), and null-point height ($z=6.13\ \mathrm{Mm}$). The configuration of case G6, with its relatively low mass density and low temperature (compared to the upper corona) and the given magnetic field strength, resembles that of quiet-Sun coronal holes \citep{2019A&A...629A..22H}. Case G6b, with its substantially higher mass density, represents plasma in the lower corona (but not in coronal holes) with a magnetic field strength similar to that of coronal loops \citep{2001A&A...372L..53N, 2008A&A...487L..17V}. 

The boundary conditions of both cases are as given in \citetalias{\secondpaperref}, with a velocity driver applied on the lower boundary.
This was done by setting the horizontal velocity component at the bottom boundary to 
\begin{align}
    u_x(x,z=0,t) = v_\mathrm{d}(t) v_0(x) \label{eq:bound-ux} \ ,
\end{align}
where the spatial component $v_0(x)$ was given by
\begin{align}
    v_0(x) = \left( \frac{1+\cos\left(\pi(x-L_x)/L_x\right))}{2} \right)^{10} , \label{eq:bound-ux-spatial}
\end{align}
with $L_x = 16$ Mm, and the temporal component $v_d(t)$ was given by
\begin{align}
    v_\mathrm{d}(t) = v_\mathrm{p} \left\{ 
    \begin{array}{ll}
         \sin\left(0.5\pi t/t_\mathrm{r}\right) & t \in [0, t_\mathrm{r}] \\
         1.0 & t \in [t_\mathrm{r}, t_\mathrm{d}-t_\mathrm{r}] \\
        \sin\left(0.5\pi(t_\mathrm{d}-t)/t_\mathrm{r}\right) & t \in [t_\mathrm{d} - t_\mathrm{r}, t_\mathrm{d}] 
    \end{array}
    \right. , \label{eq:bound-ux-time}
\end{align}
with $v_\mathrm{p} = 1\ \mathrm{km\ s^{-1}}$, $t_\mathrm{r} = 10$ min, and $t_\mathrm{d} = 40$ min. This ensured that outer spine would be moved horizontally with a peak velocity of $1\ \mathrm{km\ s^{-1}}$, thus resembling an ad hoc mechanism with a granular-like velocity.

The experiments span a domain of $32\times 32\ \mathrm{Mm^2}$ discretized over $2048 \times 2048$ grid points and were run for 40 min. 
The size of the domain was chosen in order to fit a typical fan-spine topology with a null-point at coronal heights similar to those extrapolated from observed CBPs \citep{2017A&A...606A..46G}.
In fact, the uniform temperature, density, and pressure in the reconnection site is similar to the scenario obtained in more realistic radiative-MHD simulations at the nullpoints of CBPs, such as the 2D and 3D experiments by \citet{2022ApJ...935L..21N} and \citet{2023ApJ...958L..38N}, respectively.
In our simulations, we found that the density scale height due to gravity would be $\sim 30$ Mm, that is, nearly equal to the vertical size of our domain, and since the current sheet in our simulations covers only a fraction of that vertical size, this quite large scale height justifies the neglecting of gravity in our idealized setup. Previous studies have also shown that gravity has a negligible effect on plasmoid dynamics in the corona, due to the dominating Lorentz forces in the current sheets  \citep{2020ApJ...898...90Z}. The $z$-axis is still referred to as the height axis of our domain, despite the absence of gravity. The reason for this is the fact that our simulations have a velocity driver imposed at the bottom of the fan-spine structure, thus mimicking a solar scenario where a fan-spine topology with a null-point at coronal heights is rooted in the photosphere.

The resistivity in our simulations is given by the hyper-resistivity model 
\citep{1995NordlundGalsgaard, 2011A&A...531A.154G} of Bifrost, that is, a tensor expression that scales strongly with the gradients in the magnetic field and the fluid velocity. In \citetalias{\secondpaperref} (in which the explicit mathematical definition of the hyper-resistivity term is given), we showed that this resistivity model can be applied on a moderate numerical resolution to satisfactorily mimic the plasmoid characteristics obtained with higher resolution when using uniform resistivity. In general, the simulated reconnection characteristics in terms of plasmoid sizes, peak velocities, and energy balance has been shown to depend less on resolution when applying hyper-resistivity \citep{2013PhPl...20i2109K}.

\subsection{Forward-modeling}

 Our study focuses on synthetic observables of the \Ionline{Fe}{IX}{171}, \Ionline{Fe}{X}{174}, and \Ionline{Fe}{XII}{195} lines, whose peak formation temperatures lie around 0.8 MK, 1.0 MK, and 1.5 MK, respectively. In the following, we explain the calculations of the corresponding emissivity as well as the total and specific intensities of these lines as received from the Sun without any instrumental effects. Furthermore, we describe the forward-modeling of these lines as observed with different instruments, including the currently active SDO/AIA and SO/EUI-HRI$_\mathrm{EUV}$ telescopes, along with the upcoming MUSE/SG, MUSE/CI, and Solar-C/EUVST. 

\subsubsection{Emissivity}
The emissivity of any optically thin emission line $i$ from chemical element $X$ produced in statistical equilibrium at a given temperature $T$ is given by 
\begin{align}
    \epsilon_{i} = A_X G_{i}(n_{\rm e}, T) n_{\rm e} n_{\rm H} \label{eq:emissivity} \ , 
\end{align}
where $G_{i}(n_{\rm e}, T)$ is the contribution function of the line, $A_X$ the relative abundance of the atomic element $X$ with respect to hydrogen, $n_{\rm e}$ the electron number density, and $n_{\rm H}$ the hydrogen number density. The product $n_{\rm e} n_{\rm H}$, defined as the emission measure, is already known from our simulations. To retrieve the gain function $A_X G_{i}(n_{\rm e}, T)$ for each of the above-mentioned lines, we used the CHIANTI version 10.1 database \citep[and references therein]{2023ApJS..268...52D} with abundances given in the file \verb|sun_coronal_2021_chianti.abund| provided by \citet{2023ApJS..265...11D}. 
Since the gain functions depend weakly on $n_{\rm e}$, it is common to assume a fixed electron number density for forward-modeling in the corona (in our case, $n_{\rm e} = 10^9\ \mathrm{cm^{-3}}$).

\subsubsection{Spectral synthesis: VDEM and spectral moments}

The total intensity of a line is given by the integral of the emissivity along the line of sight (LOS). The intensity as received by a telescope in a given filter which comprises the same line (along with some neighbouring lines) is similarly found by integrating the product of the temperature response function for the given filter (which depends on the gain function and the instrumental effects) and the emission measure along the LOS. Synthetic observables of spectrographs, on the other hand, are given by the specific intensity $I_{\lambda,i}$ (rather than just the total intensity). We computed this specific intensity by the following integral:
\begin{align}
    I_{\lambda,i} = \iint R_{\rm i}(T,v_{\rm LOS},\lambda) {\rm VDEM}(T,v_{\rm LOS}) {\rm d}T {\rm d} v_{\rm LOS} . \label{eq:spectra}
\end{align}
where $R_{\rm i}(T,v_{\rm LOS},\lambda)$ is the response function of line $i$ depending of temperature $T$, LOS velocity $v_{\rm LOS}$, and wavelength $\lambda$, and ${\rm VDEM}(T,v_{\rm LOS})$ is the velocity differential emission measure \citep[VDEM,][]{2019ApJ...882...13C}. 

For our forward-modeling, we applied two different LOS: the vertical $z$-direction, integrating the whole box from the top and downwards to mimic an on-disk observation; and the horizontal $y$-direction, in order to mimic an off-limb observation, assuming to that end that our computational domain has a width of 0.2~Mm,  which roughly represents the typical diameter of the larger plasmoids of our simulations. 
Response functions for the 171 \AA, 174 \AA, and 195 \AA\ lines were computed using gain functions from CHIANTI as described above, assuming Gaussian line profiles with thermal broadening (initially neglecting instrumental broadening in order to study the pure line spectra as obtained from the Sun). The VDEM was calculated with routines developed by the MUSE team \citep{2019ApJ...882...13C}, with a velocity sampling of 2.5 km s$^{-1}$ and a $\log T(\mathrm{K})$ sampling of 0.025.

With the specific intensity (of any line $i$) given by Eq.~\eqref{eq:spectra}, we compute the zeroth, first, and second spectral moments, defined as follows:
\begin{align}
    M_0 &= \int I_\lambda {\rm d} \lambda, \label{eq:mom0} \\
    M_1 &= \frac{\int \lambda I_\lambda {\rm d} \lambda}{M_0}, \label{eq:mom1} \\
    M_2 &= \frac{\int (\lambda-M_1)^2 I_\lambda {\rm d} \lambda}{M_0} \label{eq:mom2} ,
\end{align}
where the zeroth moment is the total intensity as integrated over the line profile, the first moment is the Doppler shift of the line profile, while the square root of the second moment gives the line width $\sigma$. The line width is related to the full width at half maximum (FWHM) of the line profile by $\sigma = FWHM / \sqrt{8\ln 2}$. In our maps of spectral moments without instrumental effects, the intensities are given in CGS units ($\mathrm{erg\ cm^{-2}\ sr^{-1}\ s}$), while the Doppler shifts and line widths are given in km s$^{-1}$.

\subsubsection{SDO/AIA 171 \AA}

The synthetic SDO/AIA intensity in the 171 \AA\ filter, for any chosen LOS-coordinate $s_{_{\rm LOS}}$, was estimated by the following integral:
\begin{align}
    I_{\rm SDO/AIA\ 171\ \AA} = \int {\rm d}s_{_{\rm LOS}} R_{\rm SDO/AIA\ 171\ \AA}(T) n_{\rm e} n_{\rm H} \ ,
\end{align}
where the temperature response function $R_{\rm SDO/AIA\ 171\ \AA}(T)$ was acquired using the SolarSoft \citep{1998SoPh..182..497F} routine \verb|aia_get_response|.
Since SDO/AIA is an Earth-bound instrument with a spatial pixel size of $0\farcs6$, one pixel on an AIA image covers a $\sim 0.44$ Mm wide region on the Sun. Therefore, we degraded the resolution on all our synthetic AIA images to this pixel size. We also added some Gaussian smoothing  to account for the AIA spatial resolution of $1\farcs5$, or $\sim 1.1$ Mm \citep{2012SoPh..275...17L}. Since the AIA is capable of taking images with a cadence of 12 seconds, we applied this cadence for the temporal resolution in the AIA intensity maps against time.

\subsubsection{SO/EUI-HRI 174 \AA}
For the SO/EUI-HRI 174 \AA\ filter, we used the SO/EUI-HRI 174 \AA\ response function $R_{\rm SO/EUI-HRI\ 174\ \AA}(T),$ provided to us by Dr. Frédéric Auchère, member of the Solar Orbiter team \citep[see also][]{2023arXiv230714182G}, to compute the intensity as follows:
\begin{align}
    I_{\rm SO/EUI-HRI\ 174\ \AA} = \int {\rm d}s_{_{\rm LOS}} R_{\rm SO/EUI-HRI\ 174\ \AA}(T) n_{\rm e} n_{\rm H} \ .
\end{align}
EUI-HRI$_\mathrm{EUV}$ has an spatial pixel size of $0\farcs5$ as seen from the location of the telescope. At its perihelion distance of 0.3 AU, a pixel, therefore, covers a $\sim 0.11\ \mathrm{Mm}$ wide region of the Sun, which is the pixel size we degraded our synthetic EUI-HRI$_\mathrm{EUV}$ images to. The temporal resolution is given by a cadence of 1 second.

\subsubsection{MUSE/SG \ionline{Fe}{IX}{171} and MUSE/CI 195 \AA}
Specific intensities for the MUSE/SG \ionline{Fe}{IX}{171} line were calculated by Eq.~\eqref{eq:spectra} using the same VDEMs as used for the pure line spectra (without instrumental effects), though degraded to fit a spatial pixel size of $0\farcs167\times0\farcs167$\footnote{Though the actual pixel size of MUSE/SG will be $0\farcs167\times 0\farcs4$, we only consider the smallest pixel size, only for the $xz$ field-of-view (FOV), when synthesising spectra to preserve the highest resolution achievable with MUSE/SG  \citep{2022ApJ...926...52D}.}, and using a similar response function but with an instrumental broadening in addition to thermal broadening. Hence, the total line width of any line profile seen through the MUSE/SG is given by%:
\begin{align}
    \sigma = \sqrt{\sigma_{\rm th}^2 + \sigma_{\rm instr}^2 + \sigma_{\rm n-th}^2}, \label{eq:sigma}
\end{align}
where $\sigma_{\rm th} \equiv \sqrt{k_{\rm B} T / m_{\rm Fe}}$ is the thermal width, given the iron mass $m_{\rm Fe}$, while $\sigma_{\rm instr}$ is the instrumental width and $\sigma_{\rm n-th}$ the non-thermal width. The latter quantity is defined as the observed broadening remaining after taking into account thermal broadening (which depends on the local temperature) and instrumental broadening. In our simulations, this observed extra broadening is caused by the variations in the Doppler shift along the LOS and within the pixel size. The MUSE/SG \ionline{Fe}{IX}{171} line has a spectral sampling of 14.6 m\AA\, and the instrumental FWHM is 2.9 times the spectral sampling, hence 43.34 m\AA\, or 74.25 km s$^{-1}$ in terms of Doppler velocity, found by multiplying the value in angstroms by $c/\lambda_0$, where $c$ is the speed of light, and $\lambda_0 = 171.073$ \AA. This gives an instrumental width of $\sigma_{\rm instr} = {\rm FWHM}_{\rm instr} / \sqrt{8\ln 2} = 17.98$ m\AA\ or 31.53 km s$^{-1}$ in terms of Doppler velocity. We computed the total intensity, Doppler shift, and line width from the MUSE/SG \ionline{Fe}{IX}{171} specific intensities using Eqs.~\eqref{eq:mom0}-\eqref{eq:mom2}. For the temporal resolution, we assumed a cadence of 12 seconds, which is roughly the  time needed to perform the densest raster ---requiring a reading/moving time of 0.4 s for each of the 11 slit positions--- to achieve the best spatial resolution \citep{2020ApJ...888....3D}, assuming an exposure time of 0.6 s for each slit position. Such a high cadence is only needed for images with the highest-achievable resolution in both dimensions. If high resolution is only needed in one dimension, MUSE can provide sit-and-stare images with a cadence nearly equal to the exposure time. The intensity units for the synthetic MUSE/SG spectra are given in photon count per pixel per second ($\mathrm{ph\ pix^{-1}\ s^{-1}}$), whose conversion from units of $\mathrm{erg\ cm^{-2}\ sr^{-1}\ s^{-1}}$ is as following:

\begin{align}
    %I [\mathrm{DN\ pix^{-1} s^{-1}}] = I[\mathrm{erg\ cm^{-2}\ sr^{-1}\ s^{-1}}]  \frac{\rm sr}{\rm pix} \epsilon_{\rm eff} \frac{\lambda_0}{h c} \frac{\rm e}{\rm ph} \frac{1}{G},  
    I [\mathrm{ph\ pix^{-1} s^{-1}}] = I[\mathrm{erg\ cm^{-2}\ sr^{-1}\ s^{-1}}]  \frac{\rm sr}{\rm pix} \epsilon_{\rm eff} \frac{\lambda_0}{h c}, \label{eq:erg2ph}
\end{align}
where sr/pix denotes the size of the solid angle observed by one pixel, which for MUSE/SG is $0.167 \times 0.4 \times (2\pi/360/3600)^2$. The effective area $\epsilon_{\rm eff}$ is 3.7 cm$^2$ for the 171 \AA\ filter. Furthermore, $h c / \lambda_0$ is the energy per photon. 
%The conversion factor e/ph from photon count to electron count is given by $h c / \lambda_0$ divided by 3.65 eV. The latter number is the energy required to kick out one electron from the CCD \citep{2012SoPh..275...41B}. Finally, the factor $G$ is the gain value, that is, the number of electrons per data number, which, for now, MUSE/SG assumes 16. With this, the conversion factor from photon count to data number, e/ph/$G$, is 1.24 for the 171 \AA\ line.

We also synthesized intensity maps for the MUSE/CI in the 195 \AA\ filter. For this, we computed the emissivity of the \ion{Fe}{XII} 195 \AA\ line given by Eq.~\eqref{eq:emissivity} with the gain function for the 195 \AA\ line (acquired from CHIANTI), integrating the emissivity along the LOS and degrading to MUSE/CI spatial pixel size of $0\farcs33\times0\farcs33$. Again, units were converted to $\mathrm{ph\ pix^{-1}\  s^{-1}}$ by Eq.~\eqref{eq:erg2ph}, with an effective area $\epsilon_{\rm eff}=5.0\ \mathrm{cm^2}$ for the 195 \AA\ filter. For temporal resolution, we assumed a cadence of 4 s.

\subsubsection{Solar-C/EUVST \ionline{Fe}{X}{174} and \ionline{Fe}{XII}{195}}
Solar-C/EUVST \ionline{Fe}{X}{174} and \ionline{Fe}{XII}{195} specific intensities were computed from Eq.~\eqref{eq:spectra} the same way as for MUSE/SG \ionline{Fe}{IX}{171} using similar response functions assuming the given EUVST instrumental broadening of $\mathrm{FWHM_{instr} = 40\ m\AA}$, or $\sigma_{\rm instr} = 16.99$ m\AA, and a VDEM resolution of $0\farcs16\times0\farcs16$ for preserving the highest possible resolution achievable along any chosen slit alignment. With spectral moments calculated from Eqs.~\eqref{eq:mom0}-\eqref{eq:mom2}, the intensity unit conversion into photons per pixel per second was found by using Eq.~\eqref{eq:erg2ph} with the given pixel dimensions of $0\farcs16$ along the slit and $0\farcs4$ across the slit, along with the EUVST effective areas, which for 174 \AA\ and 195 \AA\ is 0.61 cm$^2$ and 1.2 cm$^2$, respectively.

%-----------------------------------------------------------------------

\begin{figure*}
    \centering
    \includegraphics[width=\textwidth]{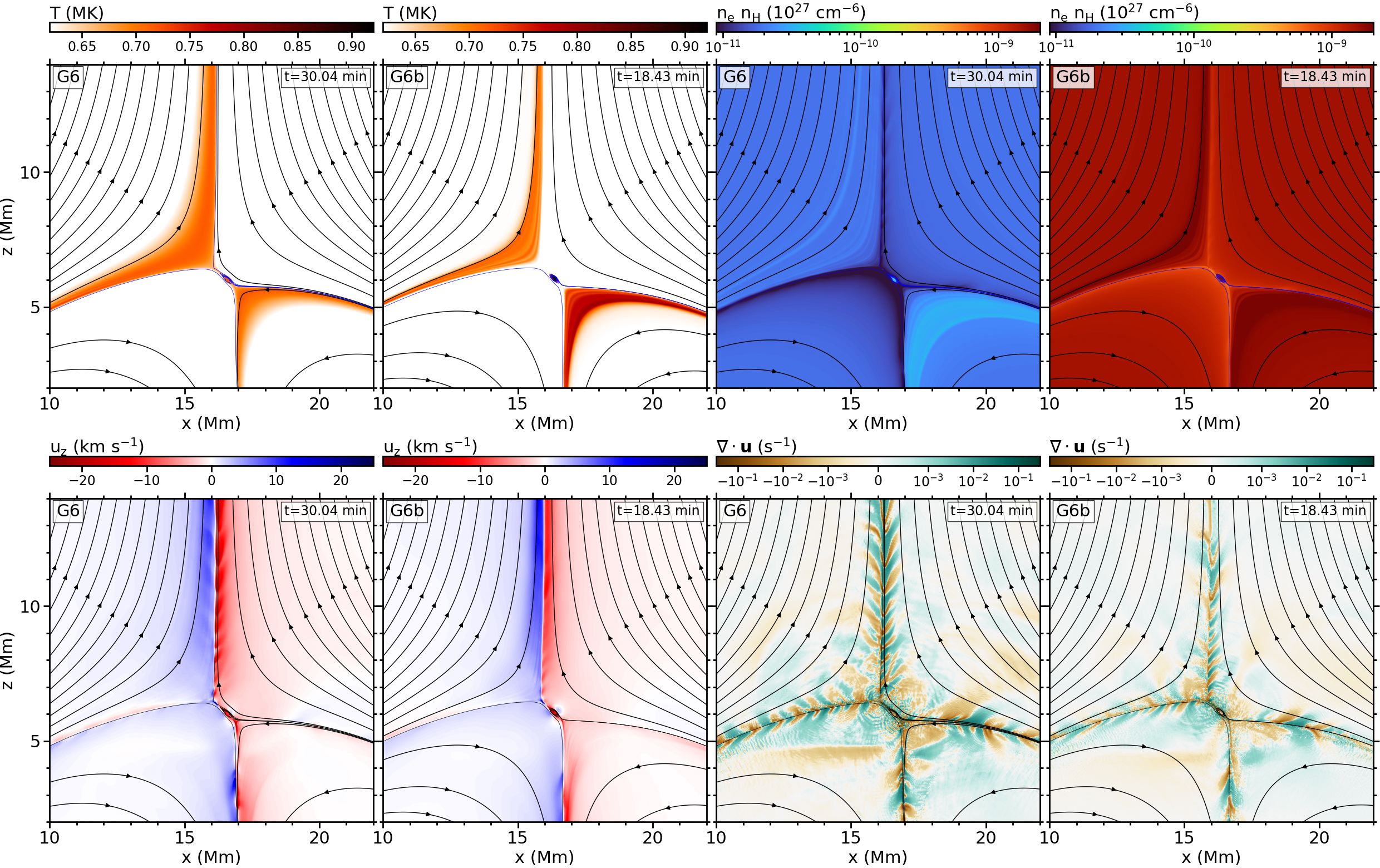}
    \caption{Temperature, $T$, emission measure, $n_{\rm e} n_{\rm H}$, vertical velocity component, $u_z$ (defined to be positive for velocities out of the Sun), and velocity divergence $\nabla \cdot {\bf u}$ with magnetic field topology superimposed in cases G6 and G6b. A movie of the full time evolution for $t\in[0,40]$ min is available online.}
    \label{fig:bfield-tg-G6B}
\end{figure*}

\section{Results}
\label{sec:results}

In the following sections, we first summarise main events of our simulations with special emphasis on the plasmoid dynamics along with the overall evolution of the temperature and velocity in the simulated fan-spine structure. Furthermore, we study each of our synthetic observables in detail in order to determine to what extent the different currently active and upcoming instruments may provide observational diagnostics on plasmoids in the solar corona. 

\subsection{Simulation summary}

Our simulations evolve similarly to the ones described in \citetalias{2024A&A...683A..95F}. 
To give a quick overview, Fig.~\ref{fig:bfield-tg-G6B} contains the temperature, $T$, emission measure, $n_{\rm e} n_{\rm H}$, vertical velocity component, $u_z$, and velocity divergence, $\nabla \cdot {\bf u}$, with the magnetic field lines superimposed for both simulations. 
Here, $u_z$ is defined to be positive for velocities pointing out from the Sun, in agreement with the $z$-axis which is defined to point outwards as well.

In each simulation, the null-point collapses as soon as the inner spine starts moving in the positive $x$-direction and a tilted current sheet is formed between the inner and outer spine. Reconnection occurs along the current sheet throughout the simulation as the inner spine keeps moving. During the reconnection process, plasmoids frequently appear along the current sheet, about 6-7 plasmoids per minute, moving either upwards to the left or downwards to the right, before submerging into either the outer spine and left fan surface or the inner spine and right fan surface, respectively. Their typical lifetimes, measured as the time from when they become visible till they submerge, are around $\sim 10-20$ s, which is shorter than the typical characteristic time scales for optically thin losses, giving a fair justification of our choice to neglect radiative cooling in our experiments, since this process would not affect the plasmoid characteristics too drastically. During these relatively short lifetimes, the plasmoids are accelerated to velocities of up to $50\ \mathrm{km\ s^{-1}}$, with equal fractions of upward- and downward-moving plasmoids. This means that they undergo an acceleration of at least $2-5\ \mathrm{km\ s^{-2}}$ due to the strong Lorentz forces of the current sheet. The gravitational acceleration, which is small ($0.274\ \mathrm{km\ s^{-2}}$) in comparison, would therefore only have a marginal effect of the plasmoid motion if included in our simulations. Occasionally, the plasmoids merge together, though coalescence instability \citep{1977PhFl...20...72F} is not a dominating part of the plasmoid dynamics in our cases due to the relatively short plasmoid lifetimes. The short current-sheet length ($\lesssim 2$ Mm) also limits the potential for plasmoid merging and most of the largest plasmoids ($\sim 0.2-0.5$ Mm) grow self-consistently without any merging and are not considered as so-called monster islands \citep[in contrast to those studied by][formed along a 4 Mm long current sheet]{2021PhPl...28i2113Z}.

The velocity divergence panels (lower right) of Fig.~\ref{fig:bfield-tg-G6B} are characterized by a wavy pattern of pulses propagating along the spines and fan surfaces, with $\nabla\cdot{\bf u}$ oscillating roughly between $-0.3$ and $0.3\ \mathrm{s^{-1}}$. These patterns are characteristic signatures of waves driven by null-point reconnection \citep{2003JGRA..108.1042G, 2017A&A...602A..43S, 2022ApJ...925..195K} and shocks triggered by the plasmoid ejection \citep{2011PhPl...18b2105Z}.\footnote{In the simulations of \citetalias{2024A&A...683A..95F}, we found that these shocks appeared in any of the simulation cases with plasmoid-mediated, or Petschek-like, reconnection, regardless of the resistivity model. The Sweet-Parker reconnection cases, on the other hand, were not characterized by such bursty shocks, but rather relatively weak and steady wave signatures.} The velocity panels (lower-left) clearly illustrate the reconnection inflows and outflows, which are particularly prominent around the spines of the structure. The temperature panels (upper-left) show that the plasma is heated due to the reconnection to temperatures up to $\sim 0.73$ MK in case G6 and $\sim 0.84$ MK in case G6b. However, the plasmoids in G6b occasionally reach temperatures close to 1 MK. The emission measure (upper right panels) is nearly two orders of magnitude higher in case G6b than in G6 due to the density difference of almost one order of magnitude. This difference has a noticeable impact on the observable traceability of the magnetic reconnection features, as seen in the following sections.

\subsection{\Ionline{Fe}{IX}{171} observables}

In this section, we study how our simulated fan-spine topology looks like in the \Ionline{Fe}{IX}{171} line, whose peak formation temperature lies within the temperature ranges of the simulated fan-spine structure. First, we analyse the pure \Ionline{Fe}{IX}{171} spectra as obtained from the Sun without any instrumental effects. Furthermore, we look at the \ionline{Fe}{IX}{171} intensity as retrieved with the currently active SDO/AIA telescope. Finally, we examine how the \ionline{Fe}{IX}{171} spectra will look like with the upcoming MUSE/SG.

\begin{figure*}
    \centering
    \includegraphics[width=\textwidth]{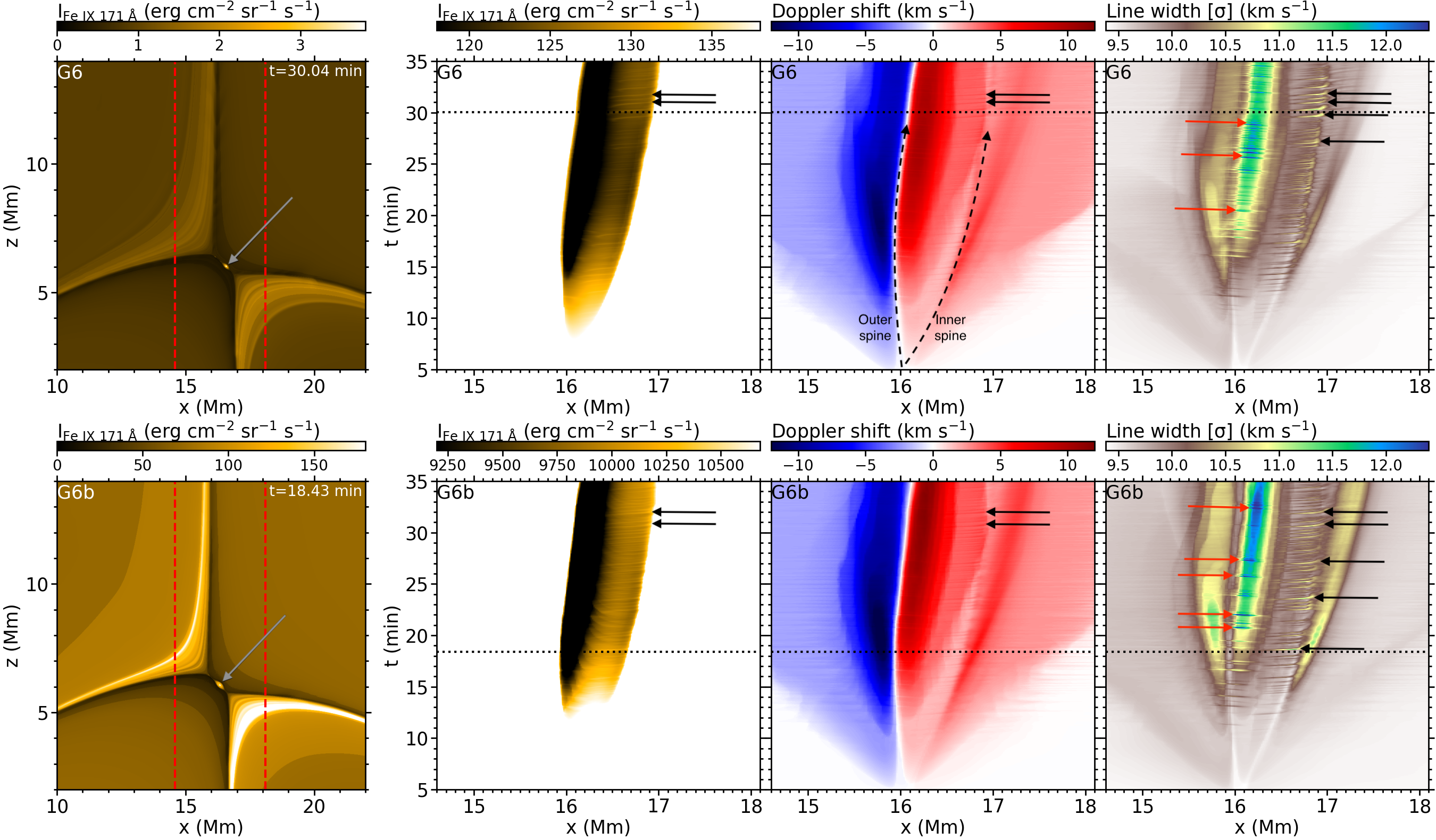}
    \caption{Synthetic spectral moments of the \Ionline{Fe}{IX}{171} line with no instrumental effects for the simulations G6 (top) and G6b (bottom). The first column contains the off-limb intensity at a given time. The other three columns show the on-disk observables for the intensity, Doppler shift, and line width, respectively, as functions of $x$ and $t$. The red dashed lines in the first column delimit the $x$ range shown in the other three columns. The black dotted lines on the latter three columns mark the time illustrated in the first column. The solid arrows point at the location of some plasmoids: red ones for those upward-moving; black or grey for downward-moving. The dashed (curved) arrows mark the inner an outer spine location (upper third panel only). A movie of the full-time evolution of the figure for $t\in[0,40]$ min is available online.}
    \label{fig:int-171}
\end{figure*}

\subsubsection{\Ionline{Fe}{IX}{171} line spectra}
\label{sec:pure171spectra}
Figure~\ref{fig:int-171} and associated animation show the synthetic \Ionline{Fe}{IX}{171} line spectra as obtained without any instrumental effects for both simulations.
The first column contains the off-limb intensity.
In both simulations, the plasmoid (marked by a grey arrow), the spines, and the fan surfaces are considerably brighter than the surrounding medium. This is due to the temperature here being closer to the peak formation temperature for this line as well as the emission measure being slightly higher than the surroundings. The overall intensity in case G6b is nearly of two orders of magnitude higher than in case G6 as a direct consequence of the similar difference in emission measure seen in Fig.~\ref{fig:bfield-tg-G6B}. 

The second column of Fig.~\ref{fig:int-171} shows the on-disk intensity.
In these maps, and all the following figures with similar maps, the range of the colorbar is intentionally set to 
highlight the plasmoid imprints, even though the intensity map gets saturated in the regions outside the current sheet. 
With this, the current sheet region is here easily located as the non-white area, being darker than the surroundings due to the slightly decreased temperature and emission measure of the above-lying plasma. Plasmoids are seen as tiny, bright, and tilted stripes and a few of them marked by arrows to ease their localization. The intensity signatures of the plasmoids approaching the outer spine appear darker than of those approaching the inner spine. This is because the region above the current sheet is darker in the areas closer to the outer spine, and the plasma above the current sheet covers a larger fraction of the LOS integration path. Therefore, the areas above the current sheet which lies closer to the outer spine obtains a considerably lower total intensity contribution than the areas further away. As a consequence, the outer-spine-bound plasmoids are in this map essentially less enlightened than the inner-spine-bound plasmoids, so that only the latter ones are seen in this map.
Again, the overall intensity is nearly of two orders of magnitude higher in G6b than in G6.

The third column of Fig.~\ref{fig:int-171} illustrates the Doppler shift of the line, also as observed on-disk. In this map, and all following Doppler shift maps, positive Doppler shift (redshift) corresponds to plasma moving  away from the observer (i.e., inwards to the Sun). With this definition, negative values of $u_z$  (downflows) contributes positively to the Doppler shift. In both simulations, the line is redshifted in the region to the right of the outer spine and blueshifted in the region to the left, in good agreement with the vertical velocity distribution mapped in Fig.~\ref{fig:bfield-tg-G6B}. The outer spine can easily be located as the distinct white, nearly-vertical stripe between the blue and red areas. The inner spine appears less distinct but can be identified as the boundary between a tilted red-white stripe and a darker red stripe. The $x$-position of both spines are marked by dashed, curved arrows (only for this figure) to ease their identification. In the current sheet region (located between the dashed arrows), plasmoids can be seen as weak, thin stripes. For the same reason as for the on-disk intensity map, we mainly see the signatures of plasmoids that move downwards (towards the inner spine), appearing as slightly darker, red stripes (see arrows). Signatures of the upward-moving, outer-spine-bound plasmoids, on the other hand, are almost not seen at all, as their intensity contribution is overshadowed by the above-lying, downward-moving plasma.

The fourth column shows the line width, $\sigma$, of the \ion{Fe}{IX} 171 \AA\ line profile. Far from the current-sheet region, where the line broadening is mainly dominated by thermal broadening, the line width is $\sim 9.5\ \mathrm{km\ s^{-1}}$, as expected, since this is approximately equal to the thermal width $\sigma_{\rm th} \equiv \sqrt{k_{\rm B}T/m_{\rm Fe}}$ for $T=0.61$ MK. Closer to the current sheet, the line width is increased to values ranging from 10 to $13\ \mathrm{km\ s^{-1}}$. The line broadening is especially high in the (green) area above the current-sheet which lies closer to the outer spine ($x\in[16.1,16.4]$ Mm for $t>20$ min), due to a strong non-thermal broadening caused by large oppositely-directed bulk velocities below and above the current sheet. In fact, non-thermal broadening is the main contributor of the additional broadening effects around the current sheet, as the thermal broadening contributions from the heated plasma is nearly negligible except for the region to the left of the outer spine and to the right of the inner spine (seen clearly as large yellow areas in G6b), where the heated plasma covers a larger LOS integration. Especially, one may note that the region $x\in[17.5,18.0]$ Mm, which lies above the strongly heated right fan-surface, has here nearly the same total line broadening as any region far from the fan-spine structure. In our simulations, non-thermal broadening is exclusively caused by variations in the LOS velocity along the LOS. Signatures of the plasmoids can be seen in this map, both close to the inner and outer spines. The upward-moving, outer-spine-bound plasmoids, marked by red arrows, appear as thin, blue stripes, and the downward-moving, inner-spine-bound plasmoids, marked by black arrows, appear as thin, yellow stripes. These stripes get thicker near the end of the current sheet, where the plasmoids collide into the spine, which also causes a further enhancement in the line broadening. The largest line width measured in the plasmoids is about $13.3\ \mathrm{km\ s^{-1}}$ in case G6b. Since the plasmoids' contribution to thermal broadening is negligible here (being smaller in width than the fan-surface, where the line width is nearly unaffected by the heated plasma), we may conclude that the visible signatures of the plasmoids here are caused by non-thermal broadening due to the motion of the plasmoids along the current sheet. The thermal width above the plasmoids is therefore still roughly around $\sigma_{\rm th} \approx 9.5\ \mathrm{km\ s^{-1}}$. Hence, since $\sigma = \sqrt{\sigma_{\rm th}^2 + \sigma_{\rm n-th}^2}$, one may estimate that the non-thermal width is around $\sigma_{\rm n-th} \approx 9.3\ \mathrm{km\ s^{-1}}$, that is, nearly-equal to the thermal width.

\begin{figure*}
    \centering
    \includegraphics[width=\textwidth]{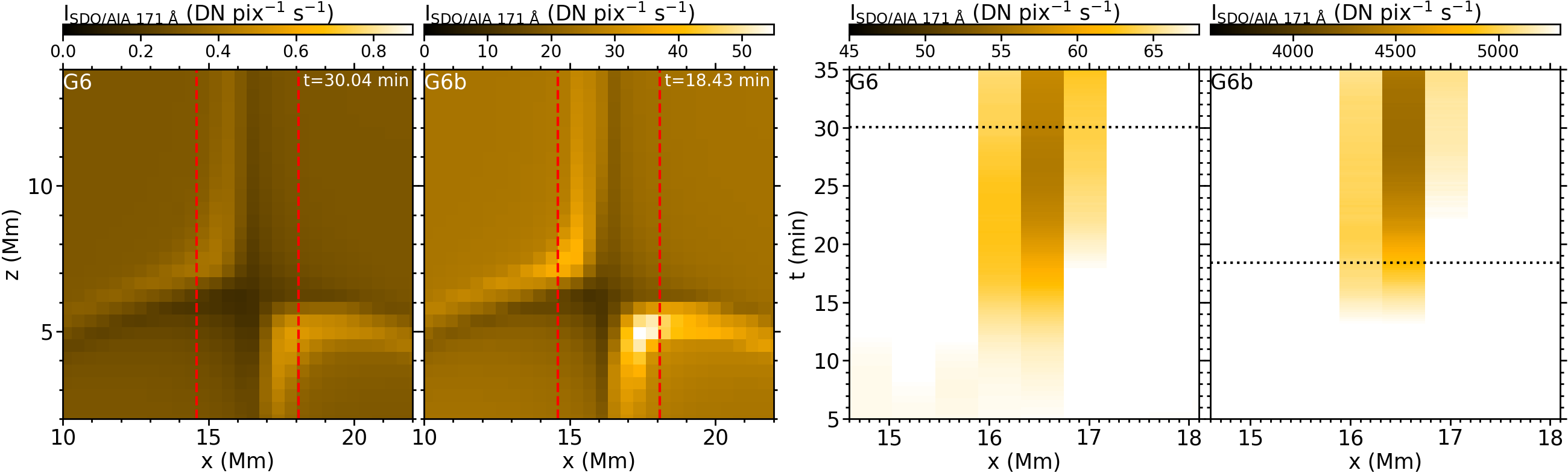}
    \caption{Synthetic SDO/AIA 171 \AA\ response for the simulations G6 and G6b as seen off-limb (two leftmost panels) and on-disk (two rightmost panels). The red dashed lines in the two leftmost panels delimit the $x$-range for the other two panels. The black dotted line in the two rightmost panels marks the time shown in the two leftmost panels.}
    \label{fig:aia-171}
\end{figure*}

\subsubsection{SDO/AIA 171 \AA\ intensities}
\label{sec:results-aia171}
The SDO/AIA 171 \AA\ intensity maps are shown in Fig.~\ref{fig:aia-171}. The two leftmost panels clearly show that the AIA instrument lacks enough resolution to resolve the plasmoids that are generated along the current sheet. The fan-spine topology is still resolvable in both cases. In G6, however, the count rate ($<1\ \mathrm{DN\ pix^{-1} s^{-1}}$) indicates that the outer spine emits barely one photon every two seconds. The photon count per pixel need to be above $\sim 100$ in order for the signal-to-noise ratio (S/N) to be satisfactorily high. It would therefore require an exposure time of at least two minutes in order for the fan-spine topology of G6 to become visible on an observational image. For case G6b, the count rate is above the desirable level for the fan-spine topology to be detectable in the off-limb view, even with an exposure time of a few seconds. As seen from the count rate in the right-side panels, the on-disk SDO/AIA 171 \AA\ response is slightly dimmer above the current sheet than in the surroundings, in agreement with what is seen in Fig.~\ref{fig:int-171}, although the current sheet here is no more than three pixels wide when viewed with AIA. The post-reconnection loops (though difficult to see in the on-disk intensity maps due to the chosen colorbar range) are still substantially brighter than the surroundings and can easily be detected with this instrument.

\begin{figure*}
    \centering
    \includegraphics[width=\textwidth]{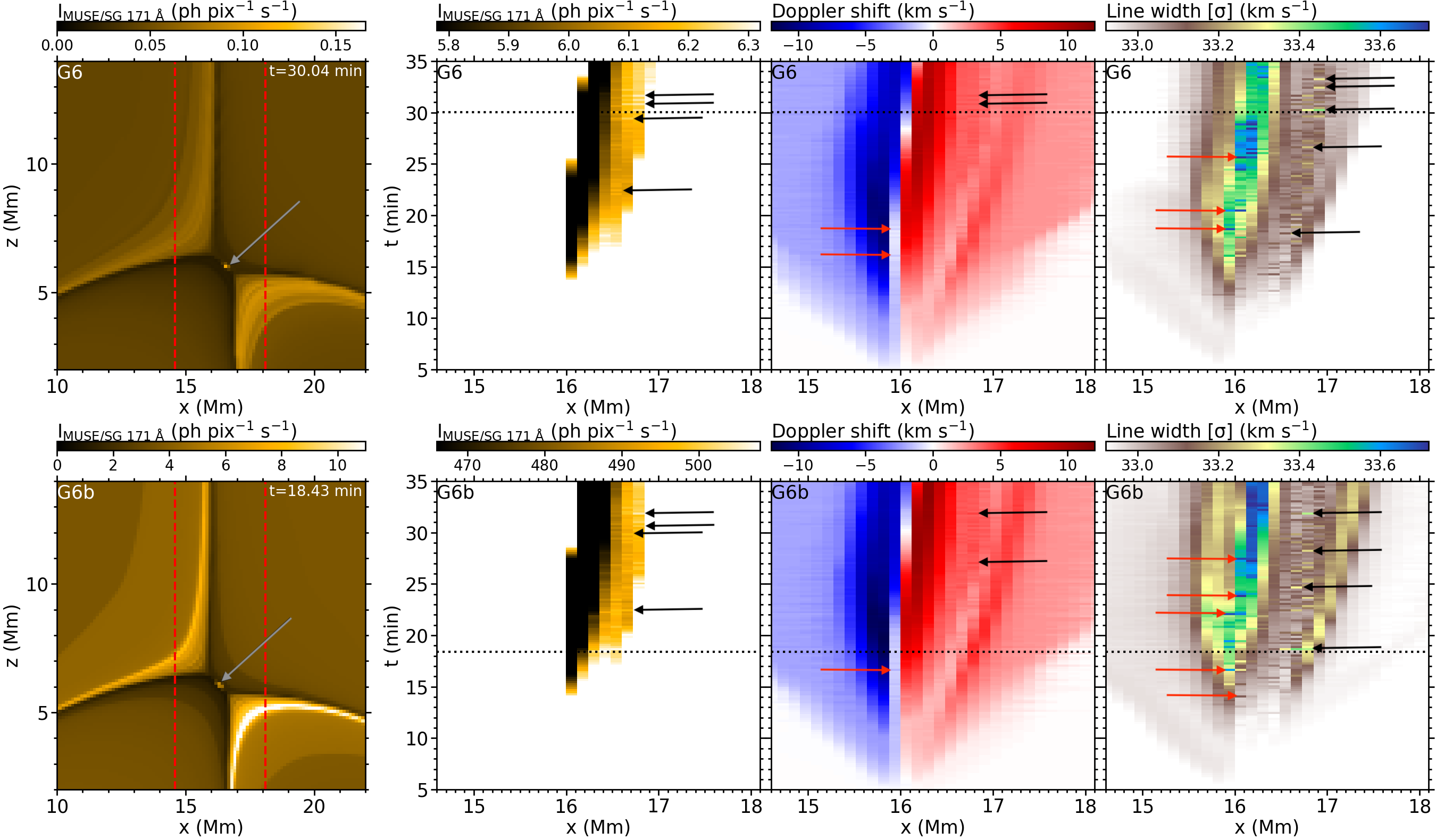}

    \caption{Synthetic MUSE/SG \ionline{Fe}{IX}{171} spectral moments for the simulations G6 (top) and G6b (bottom). Layout is the same as for Fig.~\ref{fig:int-171}. A movie of the full-time evolution of the figure for $t\in[0,40]$ min is available online.}
    \label{fig:muse-171-G6-G6b}
\end{figure*}

\subsubsection{MUSE/SG \ionline{Fe}{IX}{171} spectra}
To study the capabilities of MUSE, Fig.~\ref{fig:muse-171-G6-G6b} (and the associated movie) contains the synthetic spectral moments of the MUSE/SG \ionline{Fe}{IX}{171} line for simulations G6 and G6b. It is similar to Fig.~\ref{fig:int-171} but includes MUSE/SG  instrumental effects.
The first column shows that our plasmoids are clearly well-resolved with MUSE/SG in the off-limb view. The photon count rate for the low-density case G6, however, is below the required threshold for plasmoid detection. Even with an exposure time of 10-20 seconds, which corresponds to the typical lifetime of our simulated plasmoids, barely 1-2 photons per pixel will be received from the plasmoid, which is not enough to make the signals brighter than the noise. As for case G6b, both the fan, spines, and plasmoids have a remarkably higher photon count rate ($>5\ \mathrm{ph\ pix^{-1}\ s^{-1}}$) and can be possible to detect, for instance, by taking sit-and-stare images with the slit aligned with the current sheet and an exposure time between 10 and 20 seconds.

The on-disk MUSE/SG \ionline{Fe}{IX}{171} intensity  (second column of Fig.~\ref{fig:muse-171-G6-G6b}) is noticeably dimmer in the region above the current sheet than in the surroundings. Plasmoids are here seen as thin, white stripes (see arrows). The photon count rates indicate that plasmoids of both cases G6b and G6 should be visible through the MUSE/SG \ionline{Fe}{IX}{171} line in this on-disk view, at least with sit-and-stare images for case G6 (to allow for an exposure time of 10-20 s) and even with full-raster images for case G6b.

The corresponding Doppler shift maps (third column) look nearly the same for both simulation cases. The line is blueshifted on left side of the spine and redshifted on the right side due to plasma moving vertically in opposite directions on each side of the spine, in a similar fashion to what is seen in Fig.~\ref{fig:int-171}. Plasmoid signatures appear as thin, blue stripes (marked by red arrows), along the outer spine and as slightly darker red stripes (marked by black arrows) along the inner spine. The maximum change in Doppler shift due to the plasmoids are of order $\sim 2-3\ \mathrm{km\ s^{-1}}$.

The evolution of the line width (Fig.~\ref{fig:muse-171-G6-G6b}, fourth column) is also very similar for both simulations. The line width stays roughly around $\sim 33\ \mathrm{km\ s^{-1}}$ due to the instrumental width of $\sigma_{\rm instr}=31.5\ \mathrm{km\ s^{-1}}$, which makes the thermal and non-thermal broadening effects small in comparison. In the regions far from the spines, the plasma is at rest and not essentially heated, leaving the temperature barely above 0.61 MK. Therefore, we have $\sigma_{\rm th}=9.53\ \mathrm{km\ s^{-1}}$ and $\sigma_{\rm n-th}\approx 0$, which, together with the above-mentioned instrumental broadening, agrees with the total line width of $\sigma \approx 32.9\ \mathrm{km\ s^{-1}}$, seen in the line width maps far from the spines. At the centre of the outer spine, the line width is enhanced by $\sim 0.5-0.7\ \mathrm{km\ s^{-1}}$. Imprints of plasmoids are seen as blue stripes (marked by red arrows) along the outer spine and yellow stripes (marked by black arrows) along the inner spine. As discussed in Sect.~\ref{sec:pure171spectra}, these plasmoid signatures in the line width map are mainly due to non-thermal broadening, caused by the motion of plasmoids along the current sheet, and not by the heating of the plasmoids.  Here, the total line width reaches values up to $33.8\ \mathrm{km\ s^{-1}}$, in good agreement with the previously-estimated non-thermal broadening of about $\sim 9.3\ \mathrm{km\ s^{-1}}$. This maximum line width (seen in the plasmoids) is still only about $1 \%$ higher than the average line width measured in the same region when no plasmoids are present and, consequently, we can expect that these plasmoids will not leave any visible trace in the observed MUSE/SG \ionline{Fe}{IX}{171} line width with observational noise taken into consideration.

Synthetic Solar-C/EUVST \ionline{Fe}{IX}{171} spectra are not included here since its response function is very similar to that of MUSE/SG \ionline{Fe}{IX}{171}. With only slightly higher spatial and spectral resolution than MUSE/SG, this one-slit spectrograph can be expected to produce a nearly identical on-disk observation of the simulated fan-spine topologies of cases G6b and G6 as seen in the latter three panels of Fig.~\ref{fig:muse-171-G6-G6b}. Besides, producing a two-dimensional image mimicking the off-limb images of MUSE/SG \ionline{Fe}{IX}{171} would take longer time for Solar-C/EUVST, which only has one slit, hence achieving a considerably lower cadence.

\subsection{\Ionline{Fe}{X}{174} observables}

Similarly to the previous section, we here study synthetic observables of the \Ionline{Fe}{X}{174} line. While its peak formation temperature of 1 MK is slightly above the typical temperatures of our simulated fan-spine structure, the plasmoids of G6b occasionally reach such high temperatures. Like the previous section, we start by looking at the pure spectra without instrumental effects before delving into the \ionline{Fe}{X}{174} observables as retrieved with the following instruments: 1) the EUI-HRI$_\mathrm{EUV}$ telescope on board the recently-launched SO mission; and 2) the EUVST spectrograph on board the upcoming Solar-C mission. 

\begin{figure*}
    \centering
    \includegraphics[width=\textwidth]{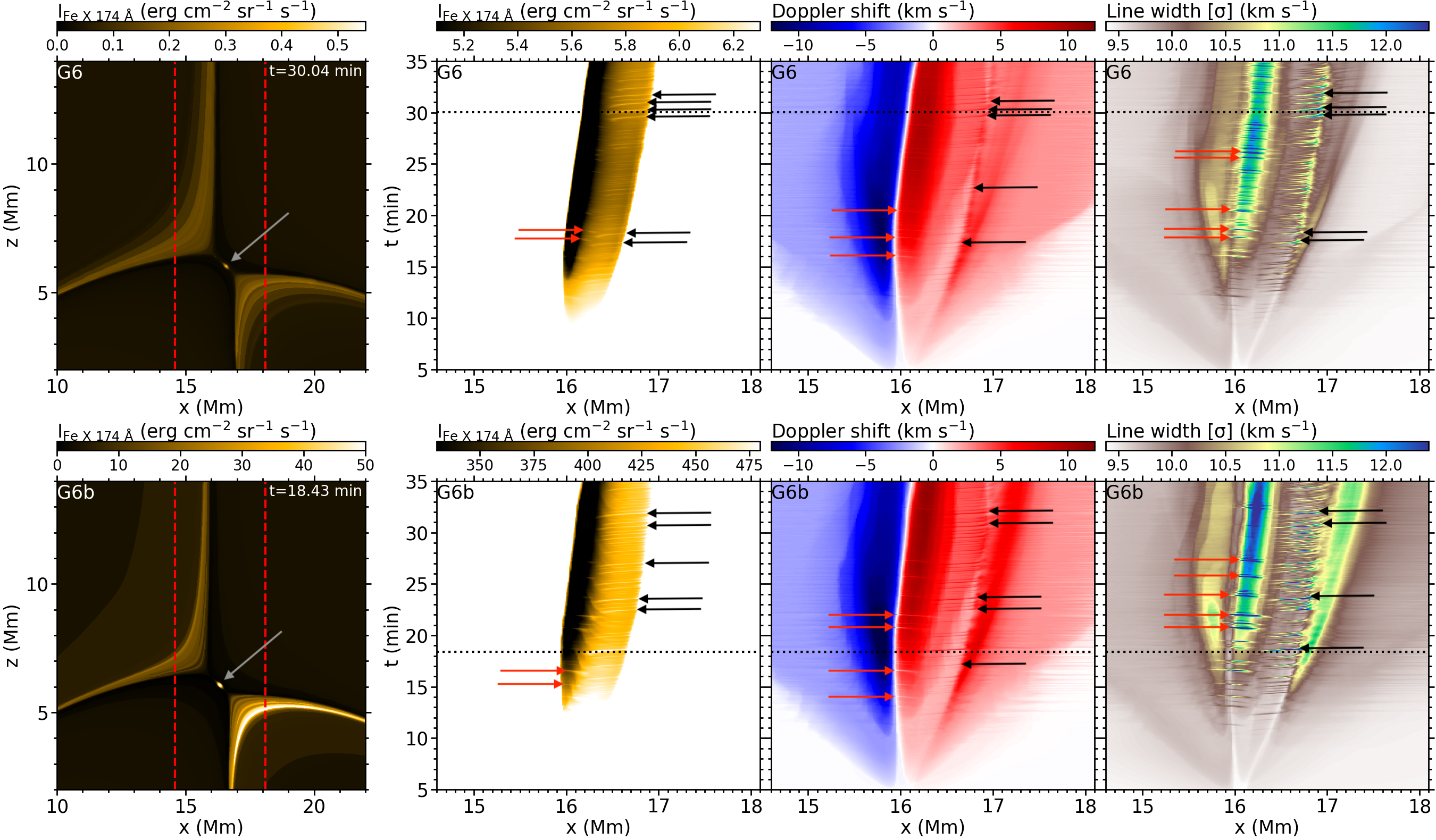}
    \caption{Synthetic spectral moments of the \Ionline{Fe}{X}{174} line with no instrumental effects for cases G6 (top) and G6b (bottom). Layout is the same as for Fig.~\ref{fig:int-171}. A movie of the full time evolution of the figure for $t\in[0,40]$ min is available online.}
    \label{fig:int-174}
\end{figure*}

\subsubsection{\Ionline{Fe}{X}{174} line spectra}

Figure~\ref{fig:int-174} and associated animation map the \Ionline{Fe}{X}{174} line spectra without instrumental effects in a similar manner as Fig.~\ref{fig:int-171} for the \Ionline{Fe}{IX}{171} line. The mapped spectral moments of the \ionline{Fe}{X}{174} line are similar to those of the \ionline{Fe}{IX}{171} line with the following differences. The fan-spine intensity is about one order of magnitude weaker here because: 1) the \ionline{Fe}{X}{174} peak formation temperature is higher than  that of \ionline{Fe}{IX}{171}, hence slightly more outside the temperature ranges  of our simulated fan-spine structures; and 2) the \ionline{Fe}{X}{174} gain function has a lower peak value as well. Furthermore, the  \ionline{Fe}{X}{174} intensity maps have larger contrasts, with the intensity being close to zero in the non-heated regions. Because of this, the plasmoids are more distinguishable in this line, making them more visible in the on-disk intensity maps (second columns) as well, as the total intensity here depends less on contributions from above the current sheet. Though the downward-moving, inner-spine-bound plasmoids (marked by black arrows) are still easiest to spot here for the same reasons as for the \ionline{Fe}{IX}{171} case, a few upward-moving, outer-spine-bound plasmoids (marked by red arrows) are visible here as well. The Doppler shift maps (third column) are very similar to those of the \ionline{Fe}{IX}{171} line, as expected, except for the imprints of the plasmoids being slightly more visible here. The inner-spine-bound plasmoids (see black arrows) are seen as dark red stripes because they move inwards, away from the observer; thus contributing in increasing the total redshift. Similarly, the outer-spine-bound plasmoids (see red arrows) are seen as light red stripes as they move outwards, towards the observer, thus contributing in decreasing the total redshift. The \ionline{Fe}{IX}{174} line width (fourth column) are also more strongly enhanced in the signatures of the plasmoids, along with the spines and fan surfaces ---compared to the \ionline{Fe}{IX}{171} line--- since the main intensity contributions for this line lie in those features (while the contributions from the surroundings are more negligible). Consequently, the imprints of the plasmoids (see arrows) are more distinct here as well. These additional line-width-increases caused by the plasmoids are still mainly from non-thermal broadening due to the motion of plasmoids along the current sheet. The regions next to the outer (blue-green) and inner (green) spines are clearly marked here by their considerably increased line widths. Hence, the low-line-width region which lies between these two regions of increased line width provides a good proxy for localising the null-point.

\begin{figure*}
    \centering
    \includegraphics[width=\textwidth]{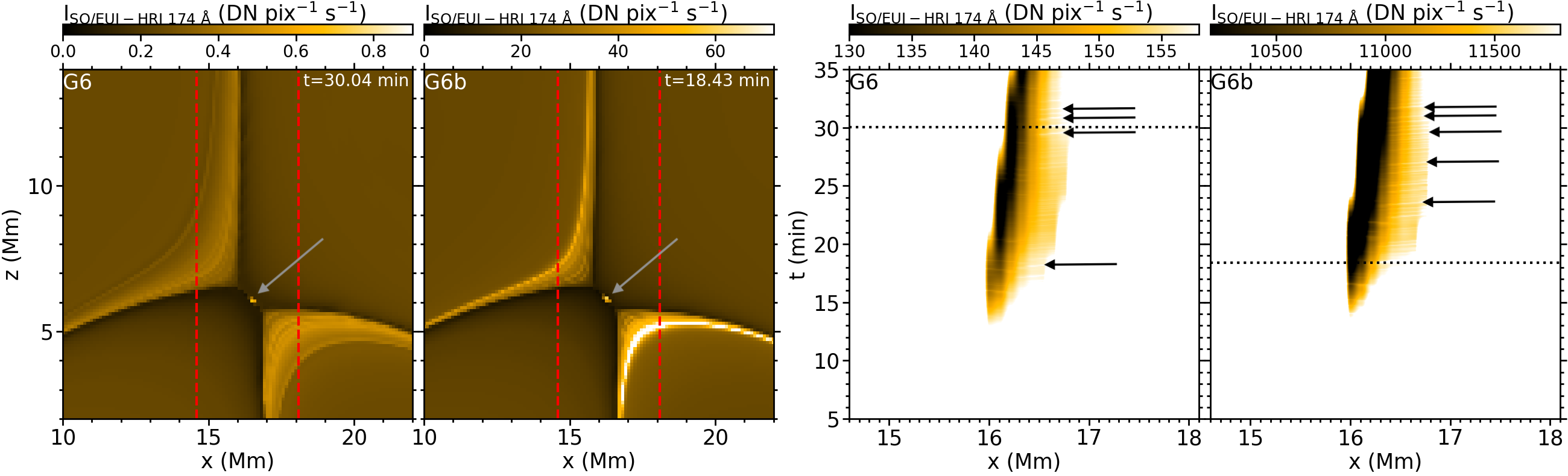}
    \caption{Synthetic SO/EUI-HRI 174 \AA\ response for the simulations G6 and G6b as seen off-limb (two leftmost panels) and on-disk (two rightmost panels). Layout is the same as for Fig.~\ref{fig:aia-171}, with the addition of arrows pointing at the location of some plasmoids.}
    \label{fig:eui-174}
\end{figure*}

\subsubsection{SO/EUI-HRI 174 \AA\ intensities}
The SO/EUI-HRI 174 \AA\ intensities are mapped in Fig.~\ref{fig:eui-174} in a similar manner as in Fig.~\ref{fig:aia-171} for SDO/AIA 171 \AA. From the two leftmost panels, we see that the contrast in the intensity maps is smaller than in the intensity maps of the isolated 174 \AA\ line. This is because the SO/EUI-HRI 174 \AA\ filter comprises several emission lines that lie near the 174 \AA\ line in the spectrum, hence its total intensity has non-negligible contributions from these lesser lines in addition to the main line. Therefore, the regions outside the fan-spine-topology are not completely dark in this filter. The count rates do actually have similar values as in the SDO/AIA 171 \AA\ intensity maps, but the resolution is remarkably better. Plasmoids are resolvable both in G6 and G6b. In the latter case, these plasmoids reach a count rate beyond the  level required for detection, given an exposure time of a few seconds. In the on-disk view (two rightmost panels), plasmoids are again seen as bright stripes in the current sheet region (see arrows) and their photon count rates are above the lower limit for detection in both cases.

\begin{figure*}
    \centering
    \includegraphics[width=\textwidth]{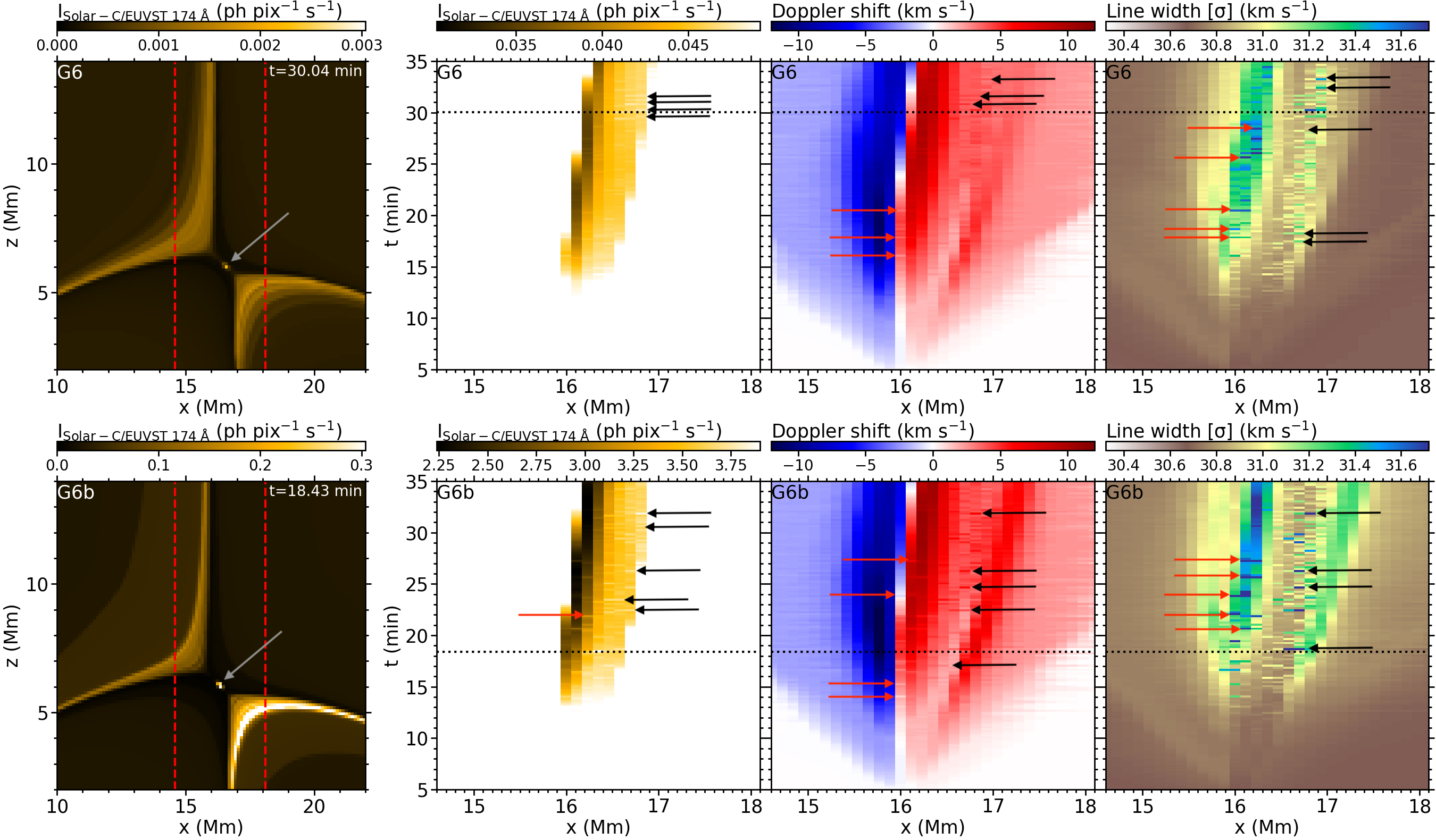}

    \caption{Synthetic Solar-C/EUVST 174 \AA\ spectral moments for cases G6 (top) and G6b (bottom). Layout is the same as for Fig.~\ref{fig:int-171}. A movie of the full-time evolution of the figure for $t\in[0,40]$ min is available online.}
    \label{fig:euvst-174-G6-G6b}
\end{figure*}

\subsubsection{Solar-C/EUVST \ionline{Fe}{X}{174} spectra}
The spectral moments of the Solar-C/EUVST \ionline{Fe}{X}{174} line are shown in Fig.~\ref{fig:euvst-174-G6-G6b} and its associated movie in a similar manner as Fig.~\ref{fig:muse-171-G6-G6b} for MUSE/SG \ionline{Fe}{IX}{171}. Even though EUVST is a single-slit spectrograph, we still choose to include a 2D map (first column) of the synthetic off-limb intensity for the sake of context and consistency with previous figures since the slit may be placed in any location above the fan-spine topology to produce a 1D projection of this map. In our cases, the photon count rates are below the desired level to detect the plasmoids in this view. The mapped on-disk spectral moments, as seen in the latter three columns, may be reproduced observationally with EUVST if the slit is aligned properly with the current sheet. In the on-disk intensity maps (second column), the plasmoids are seen as bright dots in the current sheet. The imprints of the plasmoids (see arrows) in G6b reach a photon count rate that is high enough for a detection, given an exposure time of $\sim 10-20$~s. The Doppler shift maps (third columns) show an increased redshift for the inner-spine-bound plasmoids (see black arrows), akin to the corresponding MUSE/SG \ionline{Fe}{IX}{171} spectra, and a decreased redshift for the outer-spine-bound plasmoids (see red arrows). The increases in redshift due to the downward-moving (inner-spine-bound) plasmoids seen in the Solar-C/EUVST \ionline{Fe}{X}{174} spectra are of order $\sim 5\ \mathrm{km\ s^{-1}}$, which is higher than the corresponding increases in the MUSE/SG \ionline{Fe}{IX}{171} Doppler shift and in agreement with the (non-instrumental) \ionline{Fe}{X}{174} line Doppler shift (Fig.~\ref{fig:int-174}). The decreases in redshift due to the upward-moving plasmoids are of the same order or weaker. Also, similarly to the MUSE/SG \ionline{Fe}{IX}{171} spectra, the instrumental broadening of EUVST (which is about $29.2\ \mathrm{km\ s^{-1}}$ for the \ionline{Fe}{X}{174} line) substantially overshadows the thermal and non-thermal broadening, resulting in a line width lying around $30-31\ \mathrm{km\ s^{-1}}$. Still, since the major contributions to the \ionline{Fe}{X}{174} line intensity come from the plasmoids and spines, the imprints of the plasmoids (see arrows) in the line-width maps are slightly more visible here than for MUSE/SG \ionline{Fe}{IX}{171}, with the line width being enhanced by up to 3 \% when a plasmoid hits one of the spines. In this case, the line width in the imprints of the plasmoids increases as the plasmoids hit the spines, reaching a maximum line width of $31.9\ \mathrm{km\ s^{-1}}$. The regions around the spines also have a more distinct line broadening which are directly related to thermal broadening due to the increased temperatures. This indicates that the Solar-C/EUVST \ionline{Fe}{X}{174} line may be more suitable for temperature diagnostics than MUSE/SG \ionline{Fe}{IX}{171}, especially for the regions outside the current sheet where the thermal broadening is not overshadowed by the variations in the LOS velocity along the LOS.

\subsection{\Ionline{Fe}{XII}{195} observables}

In this section, we look into synthetic observables of the \Ionline{Fe}{XII}{195} line. Despite the fact that this line has a peak formation temperature quite far above the highest plasmoid temperatures of our simulations, we still want to check for any possibility of seeing plasmoid signatures in this line, even if it should turn out that a higher temperature is needed to achieve detectability. Similarly to the previous sections, we take a look at the pure \ionline{Fe}{XII}{195} line spectra before delving into the corresponding observables seen in different instruments, in this case the upcoming MUSE/CI and Solar-C/EUVST.

\begin{figure*}
    \centering
    \includegraphics[width=\textwidth]{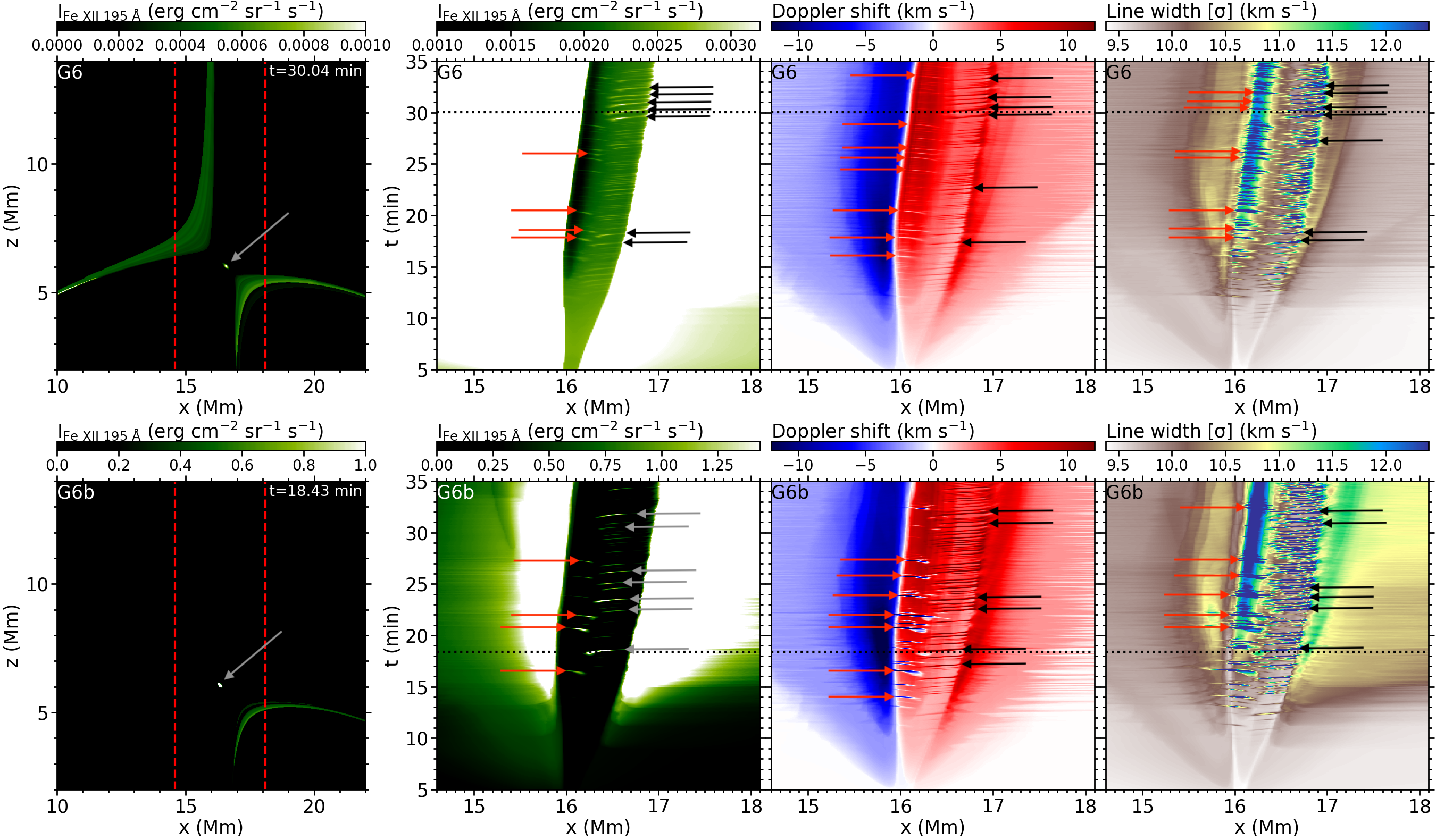}
    \caption{Synthetic spectral moments of the \Ionline{Fe}{XII}{195} line with no instrumental effects for cases G6 (top) and G6b (bottom). Layout is the same as for Fig.~\ref{fig:int-171}.  A movie of the full-time evolution of the figure for $t\in[0,40]$ min is available online.}
    \label{fig:int-195}
\end{figure*}

\subsubsection{\Ionline{Fe}{XII}{195} line spectra}

Figure ~\ref{fig:int-195} (and associated movie) shows the synthetic \Ionline{Fe}{XII}{195} line spectra without instrumental effects in a similar manner as Figs.~\ref{fig:int-171} and \ref{fig:int-174}. The mapped intensities here (first and second columns) are of several orders of magnitude lower than for \ionline{Fe}{IX}{171}, because the \ionline{Fe}{XII}{195} peak formation temperature lies at $\sim 1.5$ MK, which is far outside the temperature ranges of our simulations. The brightness contrast between the plasmoids and surrounding matter is larger, however, making the plasmoids here more distinguishable from the surroundings in the on-disk view (second columns), where we can now easily see the plasmoids moving in both directions (see red and black/grey arrows). Furthermore, since the plasmoids are the dominant components here, they are also more distinct in the Doppler shift maps (third columns). Especially in the G6b case, the upward-moving, outer-spine-bound plasmoids (see red arrows) appear as strong blueshifts (in contrast to the surrounding redshifts) instead of just decreased redshifts. The redshift enhancements caused by the downward-moving, inner-spine-bound plasmoids (see black arrows) are also stronger here than in the corresponding maps for \ionline{Fe}{IX}{171} and \ionline{Fe}{X}{174}. Finally, the plasmoids here also have a considerably higher impact on the line width (fourth column). This is because the heated plasma in the plasmoids here contributes more to the total thermal broadening, since the major contributions to the line intensity come from the plasmoids, along with the spines and fan-surfaces. For the same reasons, the line-width in the region above the fan surfaces is also noticeably enhanced, especially above the right fan-surface where the highest temperatures are reached (seen as a large yellow area in G6b).

\begin{figure*}
    \centering
    \includegraphics[width=\textwidth]{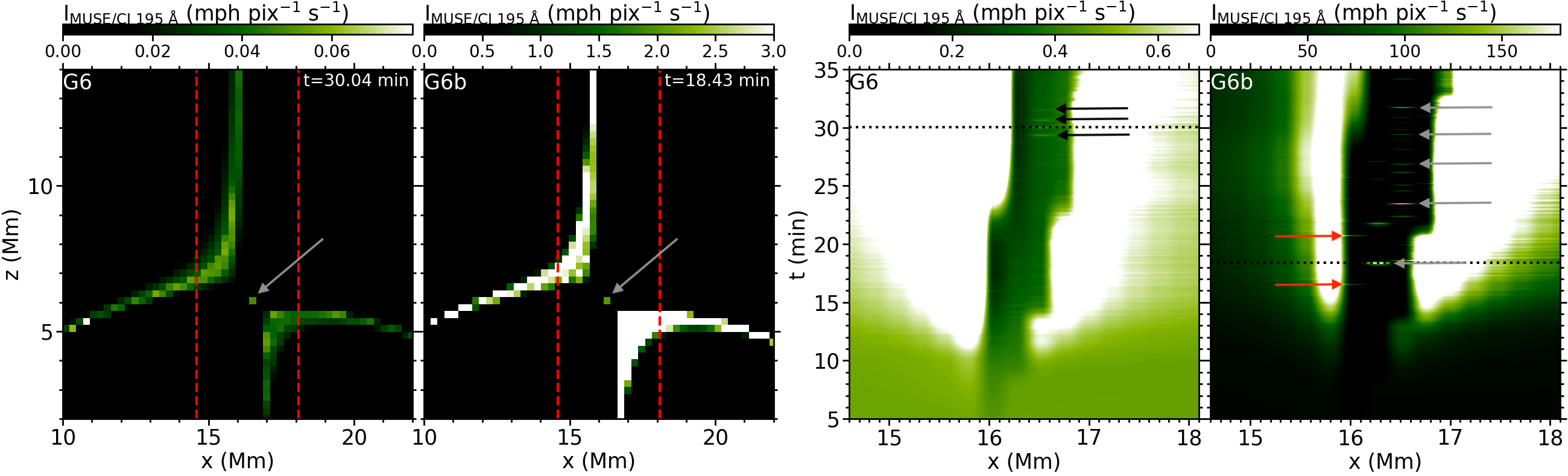}
    \caption{Synthetic MUSE/CI 195 \AA\ response for the simulations G6 and G6b as seen off-limb (two leftmost panels) and on-disk (two rightmost panels). Layout is the same as for Fig.~\ref{fig:aia-171}, with the addition of arrows pointing at the location of some plasmoids: red ones for those upward-moving; black or grey for downward-moving.}
    \label{fig:muse-ci-195}
\end{figure*}

\subsubsection{MUSE/CI 195 \AA\ intensities}
Synthetic MUSE/CI 195 \AA\ intensities are shown in Fig.~\ref{fig:muse-ci-195} in a similar manner as in Figs.~\ref{fig:aia-171} and \ref{fig:eui-174}. All four panels show that plasmoids are resolvable with the MUSE/CI (in the 195 \AA\ filter) both in the on-disk view and the off-limb view. In our cases, far below the peak formation temperature of this line, the correspondingly low photon count rates do not allow for any detection at all. The plasmoids would need to be heated up to temperatures closer to the 195 \AA\ peak formation temperature (1.5 MK) in order to be detectable, and even then, only the G6b on-disk plasmoids (fourth column) would probably be possible to detect, given an exposure time of $\sim 10\ \mathrm{s}$.

\begin{figure*}
    \centering
    \includegraphics[width=\textwidth]{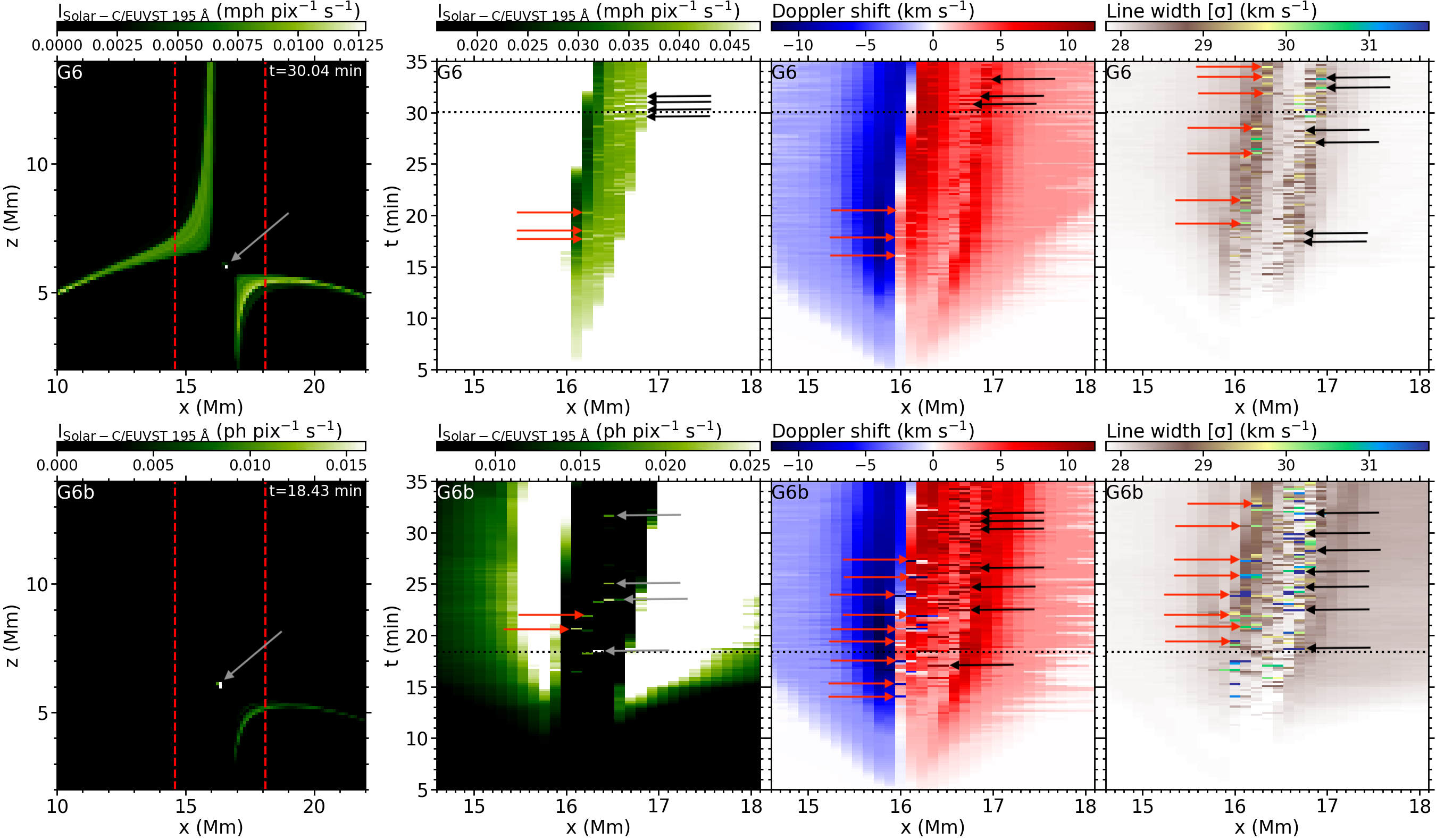}

    \caption{Synthetic EUVST 195 \AA\ spectral moments for cases G6 (top) and G6b (bottom). Layout is the same as for Fig.~\ref{fig:int-171}. A movie of the full-time evolution of the figure for $t\in[0,40]$ min is available online.}
    \label{fig:euvst-195-G6-G6b}
\end{figure*}

\subsubsection{Solar-C/EUVST \ionline{Fe}{XII}{195} spectra}
 Solar-C/EUVST \ionline{Fe}{XII}{195} spectral moments are mapped in Fig.~\ref{fig:euvst-195-G6-G6b} in a similar manner as Figs.~\ref{fig:muse-171-G6-G6b} and \ref{fig:euvst-174-G6-G6b}. The intensity maps (first and second columns) show that the plasmoids are resolvable and clearly distinguishable from the surroundings when viewing the topology both off-limb and on-disk. The photon count rates of our cases are far below the required level for detection. As already concluded from the 
 MUSE/CI 195 \AA\ intensity maps, the plasmoids would need to be heated up to temperatures closer to 1.5 MK in order to be seen in this line on-disk for the G6b case (and would still not be detectable for the G6 case). The plasmoids give stronger imprints on the Doppler shift and line width (third and fourth columns, see arrows) for this line than for the \ionline{Fe}{X}{174} line, especially for case G6b, where the plasmoids hitting the spines lead to changes of up to 10 km s$^{-1}$ in the Doppler shift and up to 3 km s$^{-1}$ in the line width. Akin to what we saw in the non-instrumental \ionline{Fe}{XII}{195} line Doppler shift map, the outer-spine-bound plasmoids leave strong bluemarks in EUVST \ionline{Fe}{XII}{195} Doppler shift maps as well, while the inner-spine-bound plasmoids are seen as strong red stripes. Especially for case G6b, a significant line broadening is seen in the region that coincides with the fan surfaces (especially the right one), indicating that this observable may potentially be used for temperature diagnostics of fan-spine topologies similar to our simulated ones.

\subsection{Spectral line profiles}

\begin{figure*}
    \sidecaption
    \includegraphics[width=12cm]{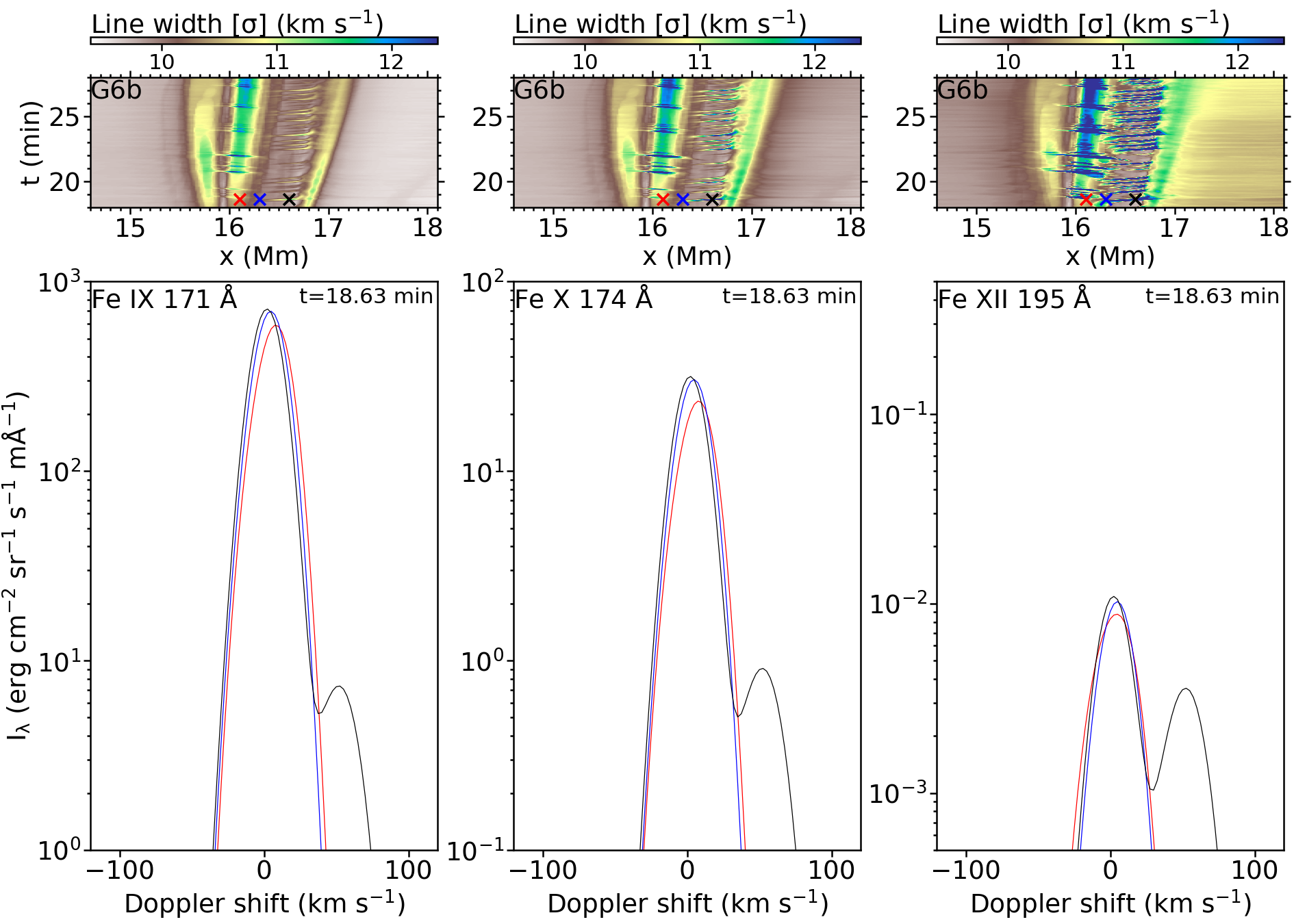}
    \caption{Spectral line profiles for the \Ionline{Fe}{IX}{171}, \Ionline{Fe}{X}{174}, and \Ionline{Fe}{X}{174} lines without instrumental effects, taken from case G6b with $z$-axis as LOS, at selected positions along $x$-axis. For context, the top panels show maps of the line width against $x$ and time, where the "x"'s mark the positions from where the corresponding line profiles with the same color (in the bottom panels) are taken. A movie of the time evolution for $t\in[18,23]$ min is available online.}
    \label{fig:spectral-line-profiles-g6b-pure}
\end{figure*}

\begin{figure*}
    \sidecaption
    \includegraphics[width=12cm]{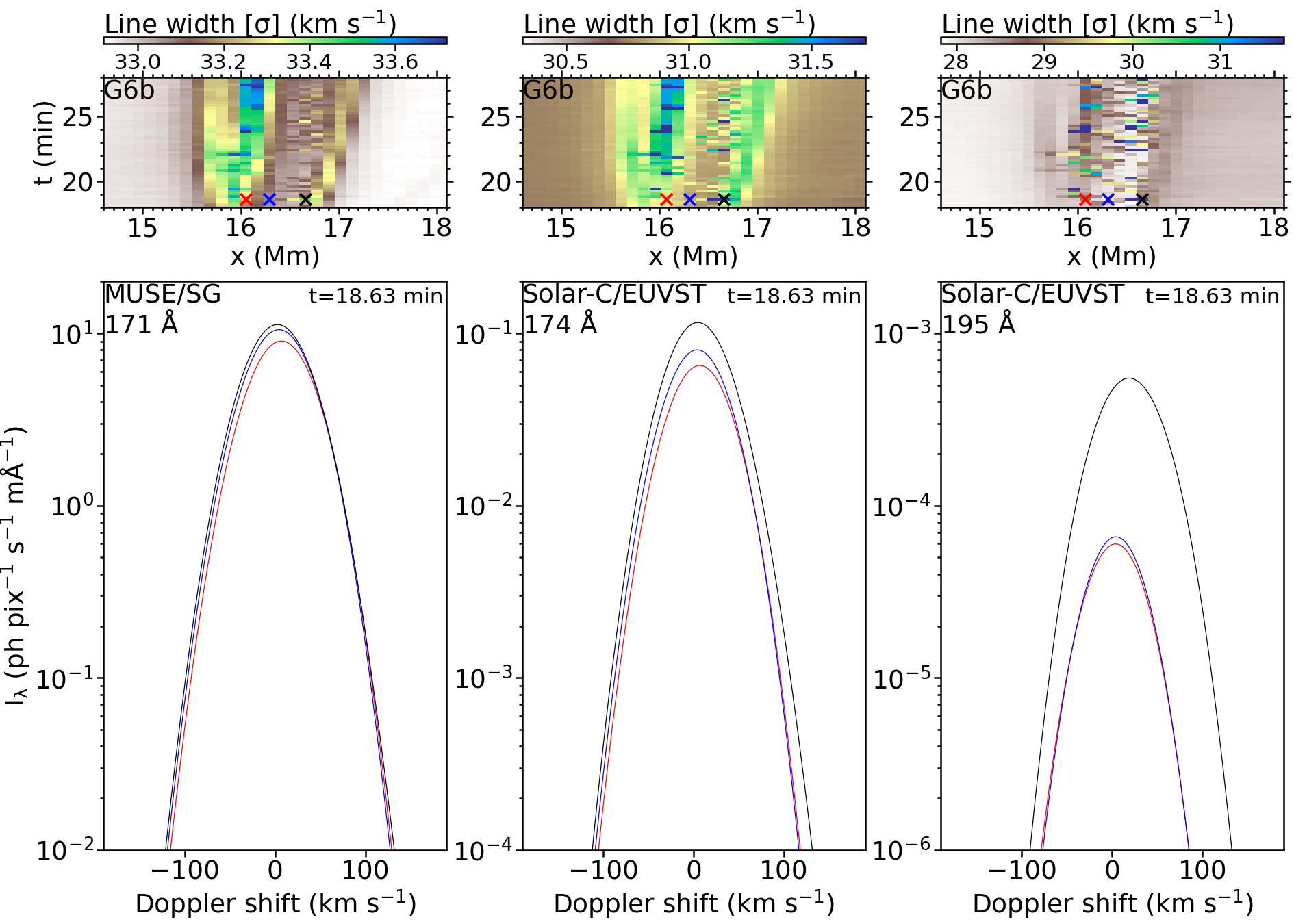}
    \caption{Spectral line profiles for the MUSE/SG \ionline{Fe}{IX}{171}, Solar-C/EUVST \ionline{Fe}{X}{174}, and Solar-C/EUVST \ionline{Fe}{XII}{195} lines, taken from case G6b. Layout is the same as for Fig.~\ref{fig:spectral-line-profiles-g6b-pure}. A movie of the time evolution for $t\in[18,28]$ min is available online.}
    \label{fig:spectral-line-profiles-g6b-MUSE-EUVST}
\end{figure*}

To complete our study, we shifted our focus to examining the synthetic spectral profiles from the G6b simulation: the one with a larger emission measure. Specifically, the top panels of Fig.~\ref{fig:spectral-line-profiles-g6b-pure} show the line widths (for context purposes) for the \Ionline{Fe}{IX}{171}, \Ionline{Fe}{X}{174}, and \Ionline{Fe}{X}{174} lines, from left to right. The bottom panels contain the synthetic profiles as observed on-disk for the locations marked in the top panels, using the same color coding. In particular, the red line profile is extracted from a location near the outer spine, the black curve is obtained close to the inner spine, and the blue curve represents an intermediate position. An associated animation of the figure is available online. 

The line profiles exhibit a nearly-perfect Gaussian shape when plasmoids are not present. However, when plasmoids pass through the marked locations, secondary peaks emerge in the line profile. This is due to the distinct vertical velocities of plasmoids compared to the surrounding plasma, which contributes to the primary peak. In the black curve, the secondary peaks appear redshifted regarding the primary peak, as the inner-spine plasmoids move downwards. Similarly, the secondary peaks in the red curve, which correspond to outer-spine-bound plasmoids, appear blueshifted. The Doppler shift of the secondary peaks reach values up to $50\ \mathrm{km\ s^{-1}}$, which is a consequence of the  plasmoids being accelerated up to such high velocities. The amplitude of the secondary peaks is typically 2-3 orders of magnitude lower than the primary peaks for the \ionline{Fe}{IX}{171} line and 1-2 orders of magnitude lower in the \ionline{Fe}{X}{174} line. Therefore, these secondary peaks will be extremely difficult to detect for those lines. With respect to the \ionline{Fe}{XII}{195} line profile, on the other hand, the secondary peaks caused by the plasmoids get comparable in size with the primary peaks, in some occasions larger. The challenge here, though, is the fact that the peak intensity is very low, about five orders of magnitude lower than for the \ionline{Fe}{IX}{171} line. 

To analyse how the instrumental effects affect the particular spectral profiles, Figure~\ref{fig:spectral-line-profiles-g6b-MUSE-EUVST} (and associated movie) represents an equivalent figure to the previous one for MUSE/SG \ionline{Fe}{IX}{171}, Solar-C/EUVST \ionline{Fe}{X}{174}, and Solar-C/EUVST \ionline{Fe}{XII}{195}. When plasmoids are not present, the profiles also have, in general, nearly-Gaussian shapes, similar to the corresponding non-instrumental line profiles. Nevertheless, the secondary peaks caused by the plasmoid motions are here mostly hidden behind the instrumental broadening. Consequently, we only perceive one peak in each of the profiles at any time. For the cases of MUSE/SG \ionline{Fe}{IX}{171} and Solar-C/EUVST \ionline{Fe}{X}{174}, the appearance of plasmoids leads to a minor broadening of the total line profile, which results in a slight increase in the total intensity, as seen in the intensity maps shown in the second column of Figs.~\ref{fig:muse-171-G6-G6b} and \ref{fig:euvst-174-G6-G6b}, respectively. Concerning the Solar-C/EUVST \ionline{Fe}{XII}{195} line profile, the instrumental broadening also makes the secondary peak indistinguishable from the primary peak, and, interestingly, the maximum of the Gaussian profile increases essentially when plasmoids are passing by. However, given the temperature of our plasmoids, the intensity of this line profile is too faint to be detected at all.

%-----------------------------------------------------------------------

\section{Discussion}
\label{sec:discussion}

In this paper, we have performed EUV forward-modeling from numerical simulations of plasmoid-mediated reconnection in the solar corona. In particular, we have synthesized observables for SDO/AIA~171 \AA, SO/EUI-HRI~174 \AA, MUSE/SG~\ionline{Fe}{IX}{171}, MUSE/CI~195 \AA, as well as for Solar-C/EUVST~\ionline{Fe}{X}{174} and \ionline{Fe}{XII}{195} to determine the capability of currently active and upcoming observational instruments for detecting small-scale plasmoids. By employing two simulations with distinct mass density parameters, we have gained preliminary insights into the lower limits of density that still allow for plasmoid detectability.

As mentioned in Sect.~\ref{sec:simulations}, we made a few simplifications in our simulations, one of them being neglecting optically thin losses. Taking these losses would, in our cases, not change the results drastically due to the following reasons. The scaling relation between plasmoid number and Lundquist number obtained using optically thin losses and ad hoc coronal heating by \citet{2022A&A...666A..28S} and the ones reproduced in our previous paper \citepalias{\secondpaperref} without optically thin losses were very similar. Furthermore, we study plasmoids with temperatures around 1 MK, which is far above the typical temperatures where optically thin losses are effective \citep[this range is around $0.05-0.2$ MK, e.g., see][and references therein]{1978ApJ...220..643R, 2008ApJ...689..585C}. In our case, the density of the plasmoids is also small (typical coronal values) and since the optically thin losses scale quadratically with the density, the cooling efficiency is limited. Since the plasmoids in our cases live shorter than the characteristic timescales in favourable conditions for optically thin losses \citep[$\sim 20-100\ \mathrm{s}$, e.g., see][]{2018ApJ...858....8N}, the plasmoids never manage to get cooled to the temperatures below 0.2 MK, where these losses dominate and lead to catastrophic cooling from which prominences and coronal rain originate \citep[e.g., see][]{1991ApJ...378..372A, 1999ApJ...512..985A, 2004A&A...424..289M, 2010ApJ...716..154A, 2021SoPh..296..143C, 2023A&A...678A.132S}.

The plasmoids in our simulations show the following characteristics: during their typical lifetimes of $10-20\ \mathrm{s}$, they reach sizes of $0.2-0.5\ \mathrm{Mm}$, get heated up to temperatures ranging between 0.73 and 1.0~MK, and are accelerated to velocities of up to $\sim 50\ \mathrm{km\ s^{-1}}$.
As a consequence of these particular properties, a widely-used instrument such as SDO/AIA is insufficient for capturing them through EUV imaging due to its moderate resolution, although the heated plasma around the fan-spine topology may still be seen in AIA 171 \AA, as shown in Sect.~\ref{sec:results-aia171}. 
Likewise, other currently active space instruments like the Spectral Imaging of the Coronal Environment \citep[SPICE,][]{2020A&A...642A..14S}, on board the SO mission, and the EUV Imaging Spectrometer \citep[EIS,][]{2007SoPh..243...19C}, on board the Hinode satellite \citep{2007SoPh..243....3K}, also offer limited spatial resolution ($1\farcs2$ and $2\farcs0$, respectively) for the detection of plasmoids akin to those we are investigating here; thus, we neglected to perform the forward-modeling for them. In contrast, in this paper we show that SO/EUI-HRI 174 \AA\ can be a suitable option for detecting such small-scale plasmoids in EUV images. In fact, recent observations have already demonstrated the SO/EUI-HRI 174 \AA\ capacity to detect brightenings with sizes down to 0.3~Mm \citep{2021A&A...656L...4B, 2023A&A...670L...3M}.

Concerning future instruments, the MUSE and Solar-C missions provide a promising perspective for observing and analysing plasmoids in the corona through EUV spectroscopy and imaging.
The design of MUSE/SG is suitable for observing plasmoids with properties resembling those from our simulations, particularly in \ionline{Fe}{IX}{171}. With its considerably high spatial and temporal resolution, it should directly detect such plasmoids through tiny, short-lived peaks in the intensity maps. In on-disk observations with the line of sight nearly parallel to the inner and outer spines, plasmoids in coronal higher density regions (e.g., $n_{\rm e} \sim 10^9\ \mathrm{cm^{-3}}$, represented in case G6b) can be detected in full-raster images with short exposure times of $\sim 0.6\ \mathrm{s}$. In lower density regions like coronal holes (e.g., $n_{\rm e} \sim 10^8\ \mathrm{cm^{-3}}$, case G6), despite their lower photon count rates, plasmoids may still be discernible through sit-and-stare images with exposure times of $\sim 10\ \mathrm{s}$. However, for off-limb observations, the lower photon count rates indicate visibility of plasmoids only in higher density regions, requiring exposure times of $\geq 10\ \mathrm{s}$, making a full-raster image impractical due to the extended time exceeding two minutes. It is also noteworthy that our plasmoids induce short-lived fluctuations of $2-3\ \mathrm{km\ s^{-1}}$ in the Doppler shift and $0.5-0.7\ \mathrm{km\ s^{-1}}$ in the line width of the MUSE/SG \ionline{Fe}{IX}{171} spectra. This additional line broadening is due to the plasmoids moving with a velocity distinctly different from the bulk velocity of the surrounding plasma, giving rise to a secondary peak in the line profile. The secondary peak can only be distinguished from the main peak in the non-instrumental line profile, as it is overshadowed by the instrumental broadening in the MUSE/SG \ionline{Fe}{IX}{171} line profile, but still contributes in increasing the total line width.
Since the MUSE/SG's specified maximum uncertainties are $5\ \mathrm{km\ s^{-1}}$ for Doppler shift and $10\ \mathrm{km\ s^{-1}}$ for the line width \citep{2020ApJ...888....3D}, it is quite likely that these plasmoid imprints could be overshadowed by Gaussian noise. Consequently, for certain detections, plasmoids would need velocities that are at least twice as fast as in our cases.

In the case of Solar-C/EUVST, we obtain similar results to those from MUSE/SG. In particular, focusing on the \ionline{Fe}{X}{174} line, 
the largest Doppler shift caused by our plasmoids is $\sim 5\ \mathrm{km\ s^{-1}}$, which could be resolved with the EUVST instrument whose maximum uncertainty is expected to be $2\ \mathrm{km\ s^{-1}}$. However, the line width broadening we obtain is still small to discern the plasmoids separately from the background, given the expected maximum instrumental uncertainty of $4\ \mathrm{km\ s^{-1}}$ for the line width. Consequently, plasmoids would need to move at least 1.5 times faster than  those from our simulations to indicate markedly distinguishable imprints in the line width.

With respect to the diagnostics focused on \Ionline{Fe}{XII}{195}, the typical temperature of our simulated plasmoids ($\sim 0.8\ \mathrm{MK}$) is far below the peak formation temperature of this line ($\sim 1.5\ \mathrm{MK}$). Therefore, we do not obtain detectable signals for either MUSE/CI 195 \AA\ or Solar-C/EUVST \Ionline{Fe}{XII}{195}. Nonetheless, it is interesting to see that if the plasmoid temperature was around 1.5 MK, Solar-C/EUVST could reveal more distinct imprints of the plasmoids, particularly in changes of the Doppler width and line shift of up to, approximately, 10 $\mathrm{km\ s^{-1}}$ and 3 $\mathrm{km\ s^{-1}}$, respectively.

For the future, it would be interesting to expand this study into 3D to get a deeper insight on how coronal plasmoids produced under fan-spine reconnection can be studied with the above-mentioned instruments. This could be done by setting up a 3D fan-spine topology akin to that of \citet{2023ApJ...958L..38N} and impose a boundary driving velocity similar to the one applied in this paper. Plasmoids in 3D tend to have the shape of solenoidal flux tubes, or magnetic flux ropes \citep{2006ApJ...645L.161A, 2017A&A...604L...7M}. Consequently, such 3D plasmoids could possibly appear as sigmoids in the synthetic on-disk intensity maps for the instruments discussed in this paper.

It is important to point out that our studies do not outrule the possibility of observing plasmoids with SDO/AIA. Synthetic EUV observables of simulated jets triggered by flux emergence \citep{2023ApJ...947L..17L} and plasmoid-fed prominence formation \citep{2022ApJ...928...45Z} clearly show coronal plasmoids resolvable with AIA. An important difference here, however, is that the above-mentioned simulations have asymmetric reconnection, meaning
that the two inflow regions of the reconnection site are clearly different in terms of density and temperature. In typical flux emergence simulations, for instance, the dense and cool emerging dome reconnects with the rarefied hot corona. Asymmetric reconnection tends to modify the onset, scaling, and dynamics of the plasmoid instability significantly \citep[e.g., see][]{2013PhPl...20f1211M, 2015ApJ...805..134M}. In such cases, as the emerging dome expands, current sheets with lengths of several megameters typically break down in several places, which would potentially allow for more plasmoids to merge and thereby become considerably larger before being ejected out of the current sheet. Our simulations, in contrast, deal with fan-spine reconnection in a uniform environment composed of purely hot and rarefied coronal plasma, basically a symmetric case in terms of inflow properties. Consequently, the plasmoids in these cases are much more challenging to observe, as they contain rarefied plasma with a typically low coronal density. Additionally, the current-sheet length is also short compared to typical flux emergence experiments, hence reducing the possibilities of plasmoid coalescence and putting a stricter limit to the plasmoid sizes. Therefore, there could be way more smaller plasmoids than those already seen with SDO/AIA. 
Our results highlight the potential of Solar Orbiter, along with the forthcoming MUSE and Solar-C missions, to study such extreme cases with small-scale plasmoids in the solar corona. This is a crucial step in advancing our understanding of plasmoid-mediated reconnection in the solar atmosphere, complementing existing research on transition-region and chromospheric small-scale plasmoids \citep[e.g.,][]{2017ApJ...851L...6R, 2020ApJ...901..148G, 2023A&A...673A..11R}.

%____________________________________________________________________________________________________________________________________
%
% Acknowledgements
%
%____________________________________________________________________________________________________________________________________
\begin{acknowledgements}
This research has been supported by the European Research Council through the
Synergy Grant number 810218 (``The Whole Sun'', ERC-2018-SyG) and 
by the Research Council of Norway through its Centres of Excellence scheme, project
number 262622.
The simulations were performed on resources provided by  Sigma2 - the National Infrastructure for High Performance Computing and Data Storage in Norway.
Juan Martínez-Sykora gratefully acknowledges support by NASA grants 80NSSC20K1272, 80NSSC23K0093, 80NSSC21K0737, 80NSSC21K1684, and contract NNG09FA40C (IRIS) and 80GSFC21C0011 (MUSE) and NSF ANSWERS grant 2149781.
%The authors are grateful to the referee for his/her constructive comments to improve the paper.
\end{acknowledgements}

\bibliographystyle{aa}
\bibliography{Faerder2024_paper3}

\end{document}